\documentclass[final,5p,times,twocolumn]{elsarticle} 

\usepackage{graphicx}
\usepackage{amsmath}
\usepackage{mathtools}
\usepackage{amssymb}

\usepackage{lineno}

\usepackage[colorinlistoftodos]{todonotes}
\usepackage[colorlinks=true, allcolors=blue]{hyperref}

\journal{Nuclear Instruments and Methods A}

\begin{document}

\begin{frontmatter}



\title{Radiation damage of SiPMs}


\author[A]{E.~Garutti}
\author[B,C]{Yu.~Musienko} 
\address[A]{University of Hamburg, Hamburg, Germany}
\address[B]{University of Notre Dame, Notre Dame, IN 46556, USA}
\address[C]{Instutute for Nuclear Research RAS, pr. 60-letiya Oktyabrya 7a, 117312 Moscow, Russia}

\begin{abstract}
The current understanding of radiation tolerance of Silicon Photomultipliers (SiPMs) is reviewed. Radiation damage in silicon sensors is briefly introduced, surface and bulk effects are separately addressed. Results on the operation of irradiated SiPMs with X-ray, gamma, electron, proton and neutron sources are presented. The most critical effect of radiation on SiPMs is the increase of dark count rate, which makes it impossible to resolve signals generated by a single photon from the noise. Methods to characterize irradiated SiPMs after their single photo-electron resolution is lost are discussed. Due to the important similarity in the operation below the breakdown voltage, also studies on radiation damage of avalanche photo-diodes (APD) are reviewed. Finally, ideas are presented on how to approach the development of radiation hard SiPMs in the future. 
\end{abstract}


\end{frontmatter}
\section{Introduction}
\label{sec:intro}
\begin{figure}[h]
\centering
\includegraphics[width=0.48\linewidth]{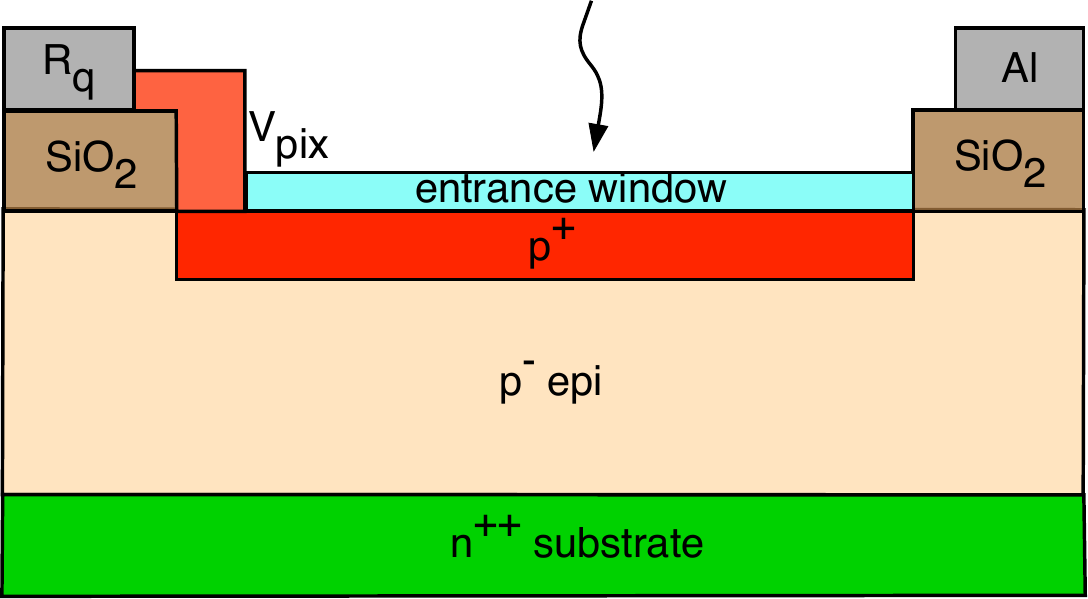}
~~~\includegraphics[width=0.44\linewidth]{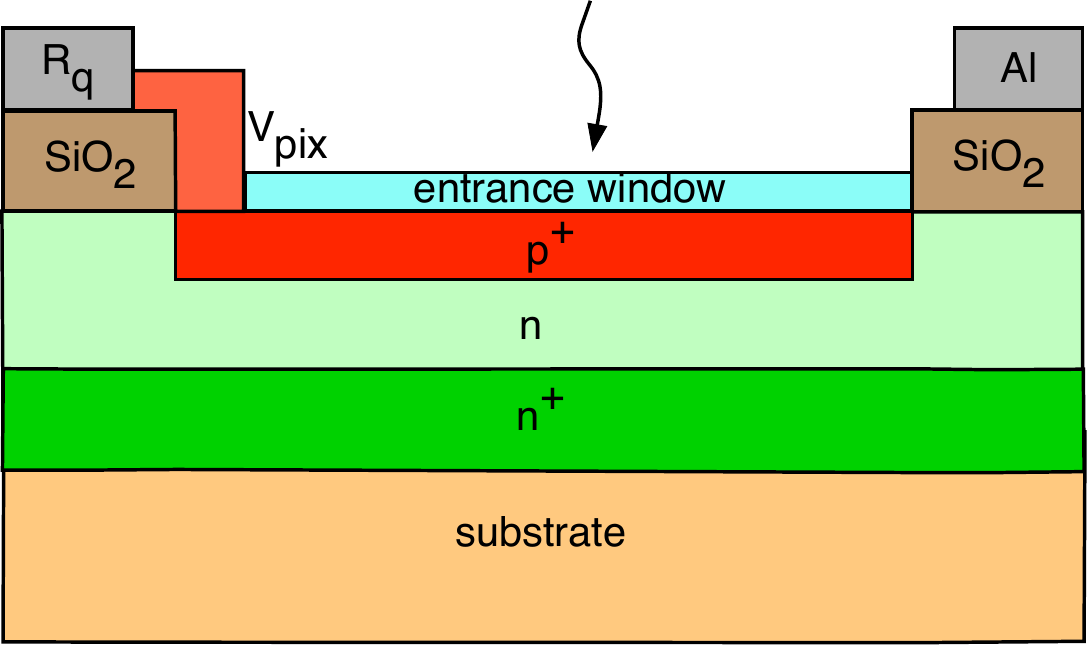}
\caption{Schematic cross-section of two possible single pixels of blue sensitive SiPMs. The poly-silicon quenching resistor $R_q$ and the Al-contact line are isolated from the silicon via a SiO$_2$ layer. A typical anti-reflecting coating material used for the entrance window  is Si$_{3}$N$_{4}$.}

\label{Fig:SiPM-crosss}
\end{figure}

This paper deals with the effects of radiation damage on SiPMs. As SiPMs detect single charge carriers, radiation damage is a major concern when 
operating these devices in harsh radiation environments 
(i.e. CMS and LHCb detectors at LHC, detectors at the proposed International Linear Collider (ILC), detectors for space experiments, etc.).  Most of the experiments at lepton colliders or at lower energy machines as well as detectors for space and medical applications will receive fluences below 10$^{12}$ particles/cm$^2$ throughout their lifetime. 
New detectors for the upgrade of the LHC experiments demand to operate SiPMs up to fluences of $\sim$10$^{14}$ particles/cm$^2$.
To understand how radiation can affect the operation of a SiPM it is necessary to first inspect its structure. 
A SiPM is a matrix of avalanche photo-diodes connected in parallel and operated above the
breakdown voltage, in Geiger mode. For a detailed description of the working principle of a SiPM we refer the reader to~\cite{Piemonte:thisVolume}.  
The single photo-diode will be referred to as pixel of the SiPM. 
Fig.~\ref{Fig:SiPM-crosss} schematically shows the cross section of two possible implementations of SiPM pixels.
The left design is used for instance in the Hamamatsu MPPC, with a depth of the p-epitaxial layer of about 2~$\mu$m, corresponding approximately to the thickness of the multiplication region. In the  right structure the p-n junction can be less deep and the multiplication region can be as thin as 1~$\mu$m or even less for special UV-sensitive designs. In both cases most of the bulk of the silicon material (denoted as "substrate") is not depleted and only a fraction of the electric charges generated in this volume will reach the multiplication region by diffusion. 
In the following we first give a short summary of the effects of radiation on the silicon crystal, including the effect on the SiO$_2$. 
Then we review radiation damage caused by electromagnetically interacting particles (photons, electrons and positrons) and by hadrons (protons, neutrons) on SiPMs in Sec.~\ref{sec:SiPM}. In Sec.~\ref{sec:apd} we discuss radiation damage of Avalanche Photo-Diodes (APD), which have been more extensively studied than SiPMs in the past and may offer important insights for future SiPM studies. Finally, we present in Sec.~\ref{sec:outlook} a summary of the factors limiting SiPM operation in high radiation environment and a list of possible precautions and design considerations when developing a SiPM for these environments. 

\section{Radiation damage of silicon - a short summary }
\label{sec:sirad}
The effect of radiation in silicon detectors below the voltage at which avalanche multiplication becomes significant is very well studied and documented in dozens of books and papers, see for reference~\cite{LUTZ1999, LINDSTROM200330}. Here we present only a short review aimed to introduce the effects relevant to SiPMs, mainly below the breakdown voltage, $V_{bd}$. 
When discussing the effect of radiation to silicon detectors one has to distinguish between two types of damage: Bulk damage due to Non Ionizing Energy Loss (NIEL), and surface damage due to Ionizing Energy Loss (IEL). 

\subsection{Radiation damage due to NIEL}
\label{sec:sirad-niel}
Bulk damage is primarily produced by high energy particles (protons, pions, electron and photons) and by neutrons, which can displace atoms out of their lattice site generating crystal defects. The minimum energy transfer required for one silicon atom displacement is $\sim$25~eV. This Primary Knock-on Atom (PKA) leaves a vacancy in the crystal lattice and relocates itself in a new position as interstitial among other atoms. If the kinetic energy of a PKA is sufficiently high ($>1$~keV) it can displace additional atoms leading to the creation of a cluster defect. For PKA energies $>12$~keV also multiple-cluster defects can form. Table~\ref{tab1} from Ref.~\cite{LUTZ1999} presents the kinematic collision properties of 1 MeV particles in Si, including Si atoms as PKA. $T_{max}$ is the maximum- and $T_{av}$ is the average transferred energy. $E_{min}$ is the minimal energy needed to create a point or cluster defect. Looking at this table, one realizes that 1~MeV electrons will produce point defects but almost no clusters, whereas protons and neutrons can produce both type of defects.

\begin{table}
\begin{tabular}{|l|l|l|l|l|}
\hline 
Radiation &  e– & p & n & Si+ \\
\hline
Interaction & EM & EM + strong & strong & EM\\
\hline
$T_{max}$&  0.155 & 133.7  & 133.9  & 1000  \\
$T_{av}$ &  0.046 & 0.210     & 50     & 0.265  \\
$E_{min}$ point   & 260  & 0.190  & 0.190  & 0.025  \\
$E_{min}$ cluster & 4600  & 15  & 15  & 2  \\
\hline
\end{tabular}
\caption{Kinematic collision properties of 1~MeV particles in Si, from \cite{LUTZ1999}. Energies are in keV.}
\label{tab1}
\end{table}

Interstitials and vacancies move inside the lattice and are very mobile above 150~K. Interstitials may annihilate with vacancies at a regular lattice position curing these defects or may diffuse out of the surface. This effect can be enhanced or sped up increasing the sensor temperature for a given time. This procedure is called annealing. Alternatively, dislocated atoms may combine with other defects and form stable secondary defects. These can be combinations of interstitials (I), vacancies (V) with C, O, P atoms, leading to permanent formation or removal of donors and acceptors, e.g.: VP, VO, Divacancy (V2), Trivacancy (V3).


According to the NIEL hypothesis the radiation damage is proportional to the non-ionizing energy loss of the penetrating particles (radiation) and this energy loss is again proportional to the energy used to dislocate lattice atoms (displacement energy).
The NIEL hypothesis does not consider atom transformations nor annealing effects and is therefore not exact.
Nevertheless, it is common to scale the damage effects of different particles using the NIEL hypothesis. 
However, different effects (leakage current, doping concentration, charge collection efficiency) require different NIEL coefficients to describe their scaling as a function of fluence.
The damage functions for various particle types and energies are discussed in Ref.~\cite{LINDSTROM200330}. It provides a hardness factor $\kappa$ for each particle type, allowing to compare the damage efficiency of radiation sources with different particles and energy spectra $\Phi(E)$. 

\begin{figure}[h]
\centering
\includegraphics[width=0.7\linewidth]{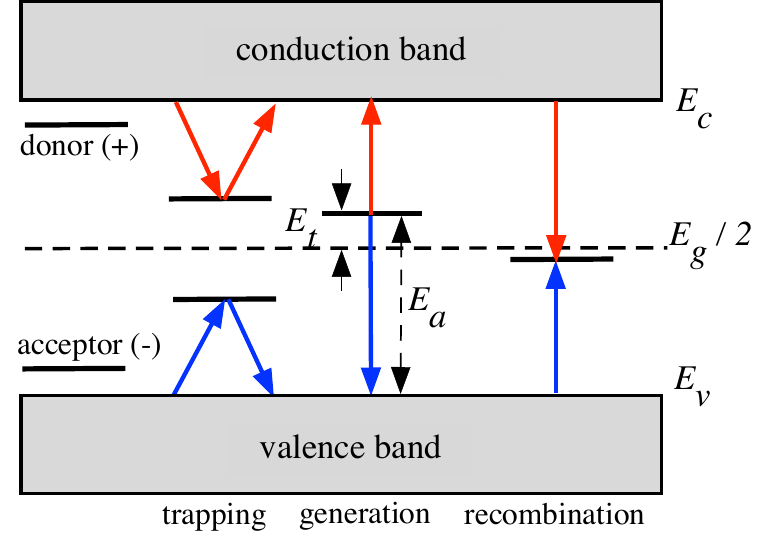}
\caption{Schematic of possible defects in the band gap of silicon and their macroscopic effect on the operation of a silicon sensor. }
\label{Fig:DefectLevels}
\end{figure}

Macroscopically, the defects generated by radiation in the silicon crystal lead to changes of the detector performance, related to the newly introduced energy levels in the energy gap between valence and conduction band, $E_g \sim$1.12 eV\footnote{This value refers to the band gap energy of silicon at room temperature.}. Depending on the position of the energy level and the capture cross sections, different effects can occur:
\begin{itemize}
\item Increase of leakage current:\\
Defects with energy levels close to the middle of the band gap ($\sim$0.56 eV) facilitate the thermal excitation of electrons and holes, increasing the dark current generated by generation-recombination. 

\item  Decrease of signal:\\
Charged defects act as trapping centers. The deep-level ones, far from the band edges have long de-trapping times and release the trapped carriers too late to contribute to the signal formation. 

\item Change of effective doping density:\\
Depending on their occupation, defect states contribute to the effective doping, and thus to the electric field in the amplification region. The occupation depends on the density of free charge carriers, on the dark current, and on the distance of the defect state from the band gap. In addition, the radiation removes dopant atoms by nuclear interactions. 
These effects have an impact on the depletion- and the breakdown voltage.


\end{itemize}
Fig.~\ref{Fig:DefectLevels} sketches the various positions of defects in the band gap of silicon and their macroscopic effects. Additionally, for devices operated in avalanche or Geiger mode, the change of the multiplication coefficient (or ionization coefficient) as function of fluence should also be considered, which is not properly discussed in the literature. 

The change of leakage current with particle fluence has been extensively studied, see for instance~\cite{Moll:1999kv} and references therein. A proportional dependence of the increase of leakage current with fluence has been reported by many authors and it is often described by the damage parameter $\alpha = \frac{\Delta I}{A\cdot w \cdot \Phi_{eq}} \sim  4 \cdot 10^{-17}$~A/cm, with $\Phi_{eq}$ being the fluence equivalent to 1 MeV neutrons, and $A\cdot w$ the depleted volume. The damage parameter $\alpha$ is often quoted at 20 $^{\circ}$C and after an annealing of 60 minutes at 80 $^{\circ}$C, for better comparison. The unwanted increase of leakage current can be mitigated by cooling the detector, or partially recovered by annealing procedures. 

To understand the temperature dependence of the leakage current it is necessary to first outline the  main contributing mechanisms: the diffusion of minority charge carriers from quasi neutral regions into the depleted region~\cite{Haitz2006,Pagano2012}; the generation of electron-hole pairs due to defects in the depletion region~\cite{LUTZ1999}, which are enhanced by a high electric field leading to the mechanism of trap-assisted tunneling or the Poole-Frenkel effect. For higher electric field strengths, direct band-to-band tunneling contributes to the generation of electron-hole pairs, as reported in~\cite{Vincent1979,Hurkx1992}. Which of the named mechanisms is dominating, depends on the electric field strength and operation temperature. The diffusion and generation currents have the following temperature dependence \cite{Grove1967}:
\begin{equation}
I_{diff} \propto T^3 e^{-\frac{E_g}{kT}},
\end{equation}
\begin{equation}
I_{gen} \propto T^2 e^{-\frac{E_a}{kT}}.
\label{Eq:Igen}
\end{equation}
For the activation energy, $E_a$ a value of 0.605~eV is found by Chilingarov Ref.~\cite{Chilin:2013}, which  using the SRH model corresponds to a trap energy $E_t= E_a + \frac{E_g}{2}$ = 45~meV from mid gap. 



A practical approximation for the temperature dependence of
the leakage current of a diode, in the absence of strong fields, i.e. amplification gain equal to one, is a decrease by a factor 2 for 8 degrees temperature reduction at room temperature. 

For electric fields of the order of $10^5$ V/cm or higher Eq.~\ref{Eq:Igen} needs to be corrected by a trap-assisted tunneling term, $I_{gen+tat}$. The correction depends on the effective field strength, $F_{eff}$ and modifies Eq.~\ref{Eq:Igen}, as
\begin{equation}
I_{gen+tat} \propto \left(1+ \Gamma\right) T^2 e^{-\frac{E_{a}}{kT}},
\label{eq:gentat}
\end{equation}
where $\Gamma \approx \frac{F_{eff}}{(kT)^{3/2}} e^{\left(\frac{F_{eff}}{(kT)^{3/2}}\right)^2} $ is the term defined by Hurkx in Ref.~\cite{Hurkx1992:field}, which accounts for the effects of tunneling.



The band-to-band tunneling current $I_{bbt}$ is independent of T, if the temperature dependence of the band-gap energy is neglected. 

For a device operating at gain larger than unity, namely in linear amplification mode (like an APD) or in Geiger mode (like a SiPM), the bulk leakage current is also linearly proportional to the amplification gain.  

\begin{figure}[h]
\centering
\includegraphics[width=0.9\linewidth]{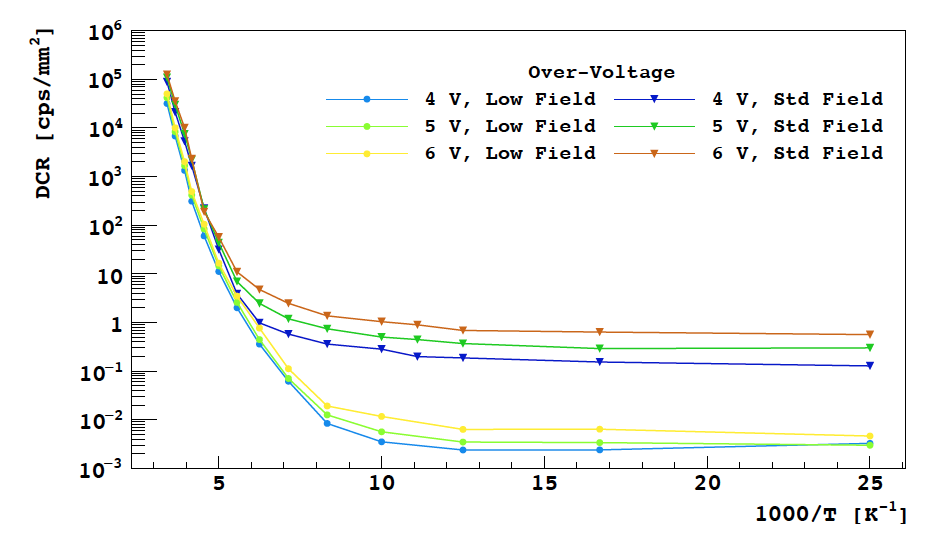}
\caption{DCR for NUV-HD FBK SiPMs with standard field (triangular markers) and low field 
(circular markers) as function of function of 1/T and over-voltage. From Ref.~\cite{Acerbi:2016ikf}.}
\label{Fig:DCRvsT}
\end{figure}

How temperature affects the operation of SiPMs is exemplary demonstrated in Fig.~\ref{Fig:DCRvsT}. 
Here, the temperature dependence of the DCR of FBK SiPMs is shown using an  
Arrhenius-like plot. Two regions are clearly separated: at high
temperatures (low $1/T$) thermal generation is the dominating mechanism responsible for the dark
rate generation. At low temperature DCR saturates and becomes T-independent. The main
contribution to DCR originates from tunneling.  Two SiPM types are presented in the plot, which differ in the design of the electric field of the multiplication region. For lower electric field value, the magnitude of the tunneling component is highly suppressed.

\subsection{Radiation damage due to IEL}
\label{sec:sirad-iel}
Surface damage is primarily produced by photons and charged particles, generating charges in the oxide (SiO$_2$) and at the Si-SiO$_2$ interface, and interface traps at the Si-SiO$_2$ interface.


Photons with energies below 300 keV, which is the threshold energy for the formation of defects in the silicon bulk, generate only surface defects. The effects of X-ray radiation damage are reported by Barnaby in Ref~\cite{Barnaby2006} and by Oldham in Ref~\cite{Oldham1999}.

In SiO$_2$, X-rays produce on average one electron-hole (e-h) pair every 18 eV of deposited energy.
Depending on ionization density and electric field, a fraction of the e-h pairs recombine. The remaining charge carriers move in the SiO$_2$ by diffusion and, if an electric field is present, by drift.
Most electrons, due to their high mobility and relatively low trapping probability, leave the SiO$_2$. However holes, which move via polaron hopping, are typically captured by deep traps in the SiO$_2$
or at the Si-SiO$_2$ interface, where interface traps are mainly formed by the depassivation of dangling bonds, resulting in fixed positive charge states and interface traps. We denote the surface density of oxide charges by $N_{ox}$, and the density of the Si-SiO$_2$ interface traps by $N_{it}$. The interface traps, if exposed to an electric field, act as generation centers for a surface current with density $J_{surf}$.

Results on $N_{ox}$ and $J_{surf}$ from MOS-Capacitors and Gate-Controlled-Diodes produced by different vendors and for different crystal orientations for X-ray doses between 10~kGy and 1~GGy are reported by Klanner~\cite{Klanner2013} and Zhang~\cite{Zhang:2013}. For a dose of 10~kGy the values for $N_{ox}$  are between 0.4$\cdot$10$^{12}$ and
1.2$\cdot$10$^{12}$~cm$^{-2}$, and for $J_{surf}$ between 0.1 and 1~$\mu$A cm$^{-2}$ at room temperature. Depending on
technology and crystal orientation for doses of the order of 1~MGy the values of $N_{ox}$ and $J_{surf}$ saturate at 1.5-3.5$\cdot$10$^{12}$~cm$^{-2}$ and 2-6 $\mu$A cm$^{-2}$, respectively. Before irradiation typical values are
a few 10$^{10}$~cm$^{-2}$ and a few nA/cm$^2$, respectively. We note that in addition to differences due to
technology, the values of $N_{ox}$ and of $J_{surf}$ at a given dose depend on the value and the orientation of the electric field in the oxide, and that there are significant annealing effects as reported by Fretwurst, Lindstroem and Moll in Refs.~\cite{Lindstrom:1999mw,Moll:1999nh}.
The depleted Si-SiO$_2$-interface areas generate surface currents, and therefore a significant increase in dark current below the breakdown voltage. If a fraction of the surface current reaches the amplification region it gets amplified and it increases also the dark-count rate. This however depends on the electric field distribution of the device. 
\\

\noindent  In conclusion, the total dark current $I_{dark}$ of a silicon device is the sum of the surface and the bulk currents. Relevant for SiPMs is to distinguish between the part of current (bulk and part of surface current) that results in a Geiger discharge and the part that does not.  


\subsection{Defect characterization}
\label{sec:defects}
Various techniques are available for the microscopic investigation of defects in silicon. Often used ones are deep-level transient spectroscopy (DLTS), thermally stimulated current (TSC) techniques and IR-absorption spectroscopy. 
\begin{figure}[h]
\centering 
\includegraphics[width=0.49\linewidth]{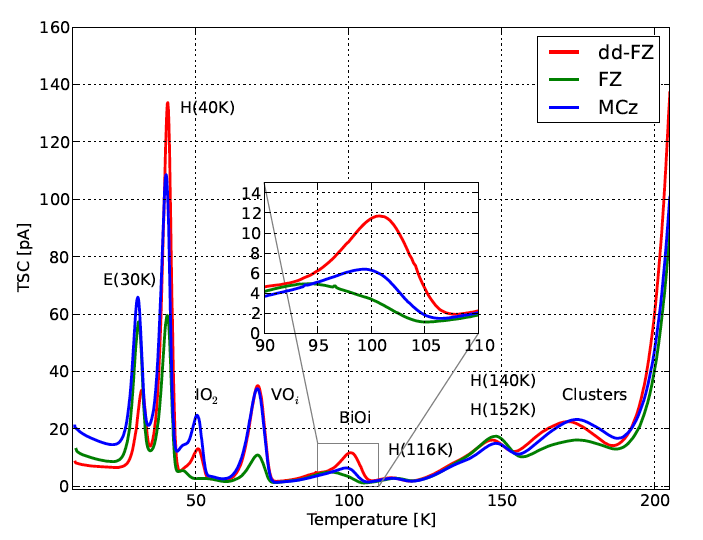}
\includegraphics[width=0.5\linewidth]{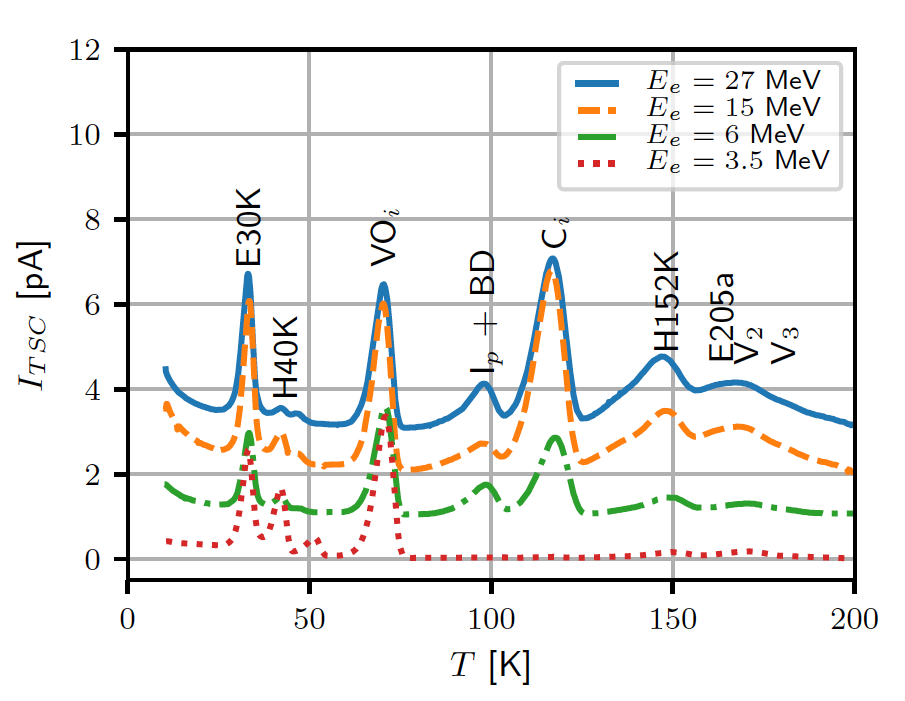}
\caption{Left) TSC spectra for 200~$\mu$m thick p-type silicon pad diodes, after 23 MeV proton irradiation ($\Phi_{eq}=0.5\cdot 10^{14}$~cm$^{−2}$) and annealed for 8 minutes at T=80 $^{\circ}$C. From Ref.~\cite{RADECSDonegani}. Right) TSC spectra,  after irradiation by electrons and normalized to $\Phi_{eq}=10^{14}$~cm$^{−2}$, and annealed for 30 min at at T=80 $^{\circ}$C. The x-axis of the plots can be converted in energy from the conduction or valence band depending on the type of defect.   High temperatures correspond to energies with large distance from the bands, i.e. close to mid gap energy.  
From Ref.~\cite{DONEGANI201815}.}
\label{Fig:RADECS}
\end{figure}
Some typical results of defect spectroscopy on irradiated silicon diodes are shown in Fig.~\ref{Fig:RADECS}. The peaks in the spectrum correspond to specific electrically charged defects in the silicon band gap. The left plot compares the defects generated by proton irradiation in silicon materials (Magnetic Czochralski and Float-Zone). The right plot shows defects generated in a silicon diode irradiated with electrons of various energies. 
In Ref.~\cite{RADECSDonegani} a correlation between the leakage current and the concentrations of three deep defects close to mid-gap (V$_2$, V$_3$ and H(220K)) is reported; and changes in the space charge in p-type sensors are correlated to the concentration of the donor E(30K), the acceptor BiOi, and of the three main deep acceptors (H(116K), H(140K) and H(152K)).

These techniques have been tested also on SiPMs.  Due to large capacitance and high dopant concentration DLTS and TSC are not applicable for SiPMs. The DLTS sensitivity limit is about $N_{trap}=10^{-4} \cdot N_d$ with $N_d$ the doping concentration. For non-irradiated SiPMs the doping concentration is $N_{d} \leq 10^{16}$ cm$^{-3}$, so one would be sensitive to a trap concentration of $N_{trap}>10^{12}$ cm$^{-3}$, which is too high with respect to the expected trap concentration from radiation damage\footnote{This holds true for low and medium fluences up to $\Phi_{eq} \sim 10^{12}$~cm$^{-2}$. For higher fluences the trap concentration may reach a measurable level. Still the very high current is a limit for DLTS application, while the very small depleted volume may be a limit for TSC applications. This still has to be verified experimentally.}. Single pixel structures yielding a smaller dark current may be better suited for this characterization, but still the high doping concentration required for the multiplication region may be an obstacle. 
IR-spectroscopy measurements on SiPMs are not reported in the literature, but could be an alternative way to characterize generation of defects by radiation damage. 

\begin{figure}[h]
\centering
\includegraphics[width=0.65\linewidth]{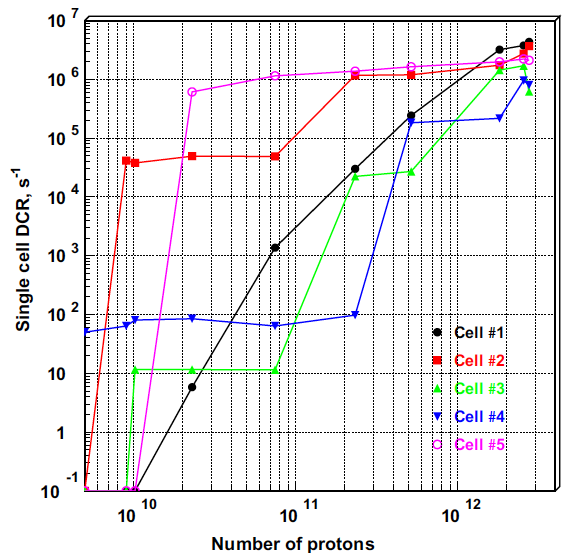}
\caption{The dark count rate of individual cells of a Philips digital photon counter as a function of total accumulated fluence. From Ref.~\cite{Barnyakov:2016bwt}. The x-axis is the total number of protons traversing the SiPM. The fluence for the highest irradiation was $\Phi_{protons} \sim 4\cdot 10^{11}$~cm$^{-2}$, and the hardness factor for this proton energy is $\kappa \sim$ 1.2.}
\label{Fig:DigiSIPM}
\end{figure}

An interesting study of the position dependence of radiation damage in SiPM pixels has been presented by  Barnyakov in Ref.~\cite{Barnyakov:2016bwt}. The authors  have investigated the radiation damage of digital SiPMs exposed to 800~MeV protons. In a digital SiPM, the DCR of every individual cell can be monitored separately, such that it is possible to generate plots like Fig.~\ref{Fig:DigiSIPM} where the DCR of single cells as function of the number of beam protons is shown.
The step-like increase of the DCR indicates that a single interaction of a proton with a  Si atom may result in a drastic DCR increase and that the increase may differ by orders of magnitude for each proton interaction. Most likely this effect is linked to the formation of cluster-like defects in one pixel. 

An elegant way to visualize the effect of radiation damage is to observe the light emission in the dark from a SiPM biased above breakdown. This is based on the phenomenon that every Geiger discharge emits a certain number of optical and IR photons produced in the high field region\footnote{The phenomenon of the emission of light with energies larger than 1.12 eV (the band gap energy in silicon) from a p-n junction operated above breakdown voltage was first observed by Newman \cite{Newman}, then quantified by Lacaita, Zappa and Bigliardi, \cite{Lacaita} to produce in average 3$\cdot$10$^{-5}$ photons per charge carrier crossing the junction. The dominant production mechanism is bremsstrahlung, but other effects also contribute.}. For randomly distributed DCR on the SiPM volume, the light emission is expected to be homogeneous. In the case of local defects in silicon, hotspots can form, which are more likely to generate Geiger avalanches in the dark.  

\begin{figure}[h]
\centering
\includegraphics[width=0.65\linewidth]{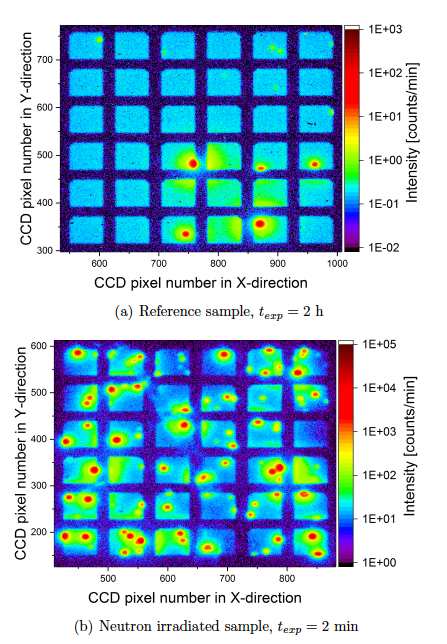}
\caption{Light intensity images for a non-irradiated (a) and a neutron irradiated SiPM test structure, operated at $\Delta$V = 4~V. The irradiated SiPM was exposed to $\Phi_{eq} = 10^{10}$ cm$^{-2}$ and no annealing was applied. The effect of radiation in increasing the number of hot-spots is evident in these images. From~\cite{Engelmann:phd}.}
\label{Fig:Engelmann}
\end{figure}

Fig.~\ref{Fig:Engelmann} from Engelmann PhD thesis, \cite{Engelmann:phd}, demonstrates very clearly the increase in the number of hot-spots due to the effects of radiation. A SiPM exposed to  $\Phi_{eq} = 10^{10}$ cm$^{-2}$ is compared to a non-irradiated device of the same type. The color scale indicates the intensity of the light emitted. 
It can be seen that the light intensity emitted\footnote{It should be remarked that the light intensity emitted by Geiger discharges is at a given gain is relatively constant, such that the measured intensity on the CCD can be used to calculate the number of discharges of the SiPM in the measurement time interval.} by the neutron induced defects (Fig.~\ref{Fig:Engelmann}b) is about two orders of magnitude larger than for ion implantation induced defects visible in the non-irradiated sample (Fig.~\ref{Fig:Engelmann}a). From this result the author concludes that
the dark count rate generated by neutron induced crystal defects is two orders of magnitude larger than the dark count rate generated by implantation defects. However, one has to consider that the implantation defects are evaluated after an annealing process, whereas the neutron defects are evaluated without a significant annealing.


\subsection{Annealing of radiation damage (defect kinetics)}
\label{sec:sirad-annealing}
\begin{figure}[h]
\centering
\includegraphics[width=0.8\linewidth]{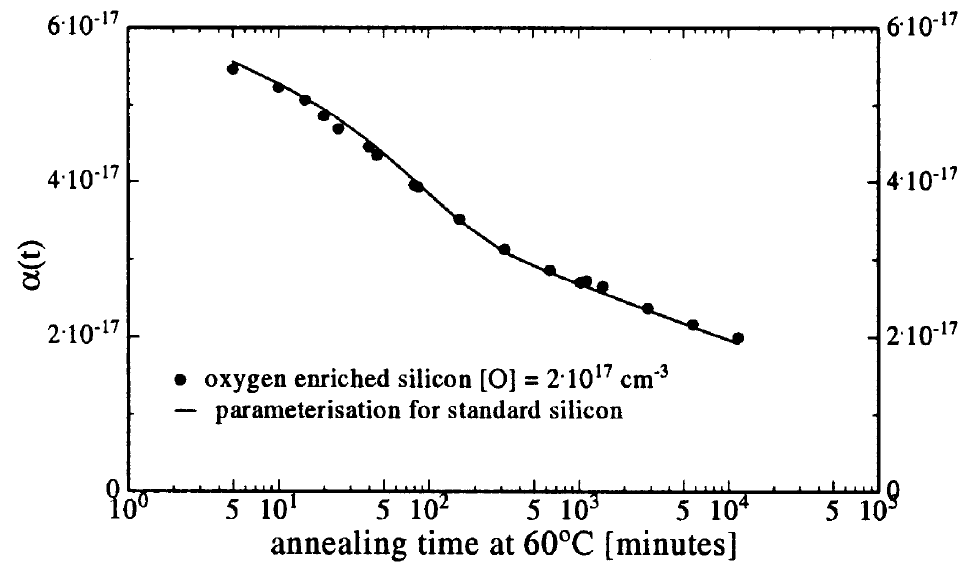}
\caption{Annealing function of current-related damage parameter
$\alpha$ for standard (solid line) and DOFZ (dots) silicon
detectors, taken from~\cite{LINDSTROM200330}.}
\label{Fig:annealing}
\end{figure}

As discussed the reverse current of a diode without gain increases proportionally to the fluence of non-ionizing radiation. To compare data, the damage parameter $\alpha$ is normalized to a given temperature (usually 20~$^{\circ}$C). It is observed in Refs.~\cite{Moll:1999kv,Moll:1999nh} that $\alpha$ can be reduced by annealing, an effect which is referred to as beneficial annealing. Fig.~\ref{Fig:annealing} demonstrates that a factor 2-3 reduction of $\alpha$, and therefore of the leakage current of an irradiated diode, is attainable after 830 hours at 60~$^{\circ}$C. Permanent damage will not be cured by this procedure such that the leakage current level of the diode before irradiation can not be fully recovered. 

A recommendation from Lindstroem~\cite{LINDSTROM200330} for the comparison of radiation damage effects of various devices is to perform an annealing of 80 min at 60~$^{\circ}$C (or equivalently 10 min at 80~$^{\circ}$C) to have results with little dependence on the detailed temperature history after the irradiation. 




\section{Radiation damage of Silicon Photomultipliers}
\label{sec:SiPM}
\begin{figure}[h]
\centering
\includegraphics[width=0.90\linewidth]{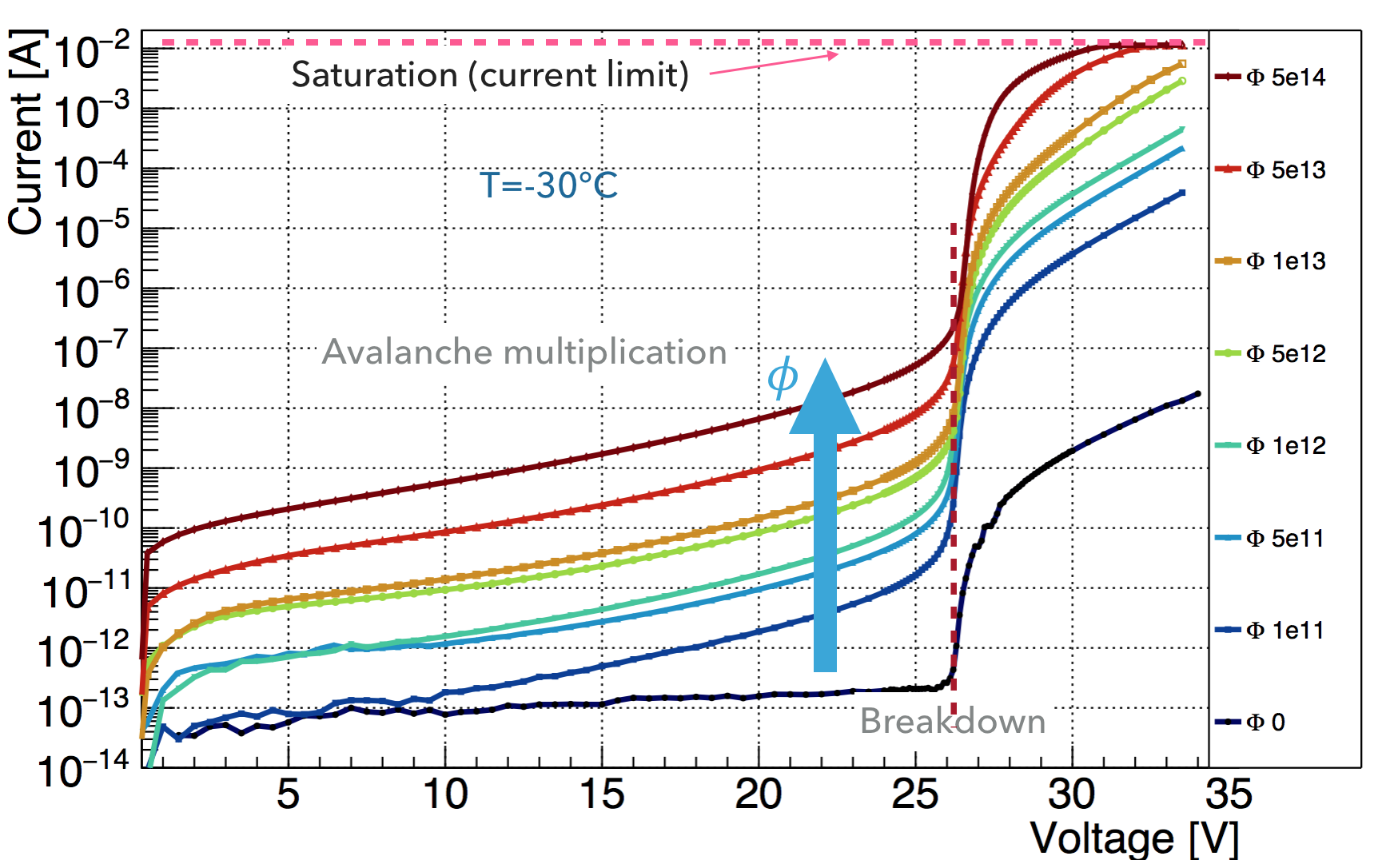}
\caption{Exemplar current-voltage curves for a KETEK SiPM (15 $\mu$m pixel size) irradiated with neutrons up to $\Phi_{eq} = 5 \cdot 10^{14}$~cm$^{-2}$ and operated at -30~$^{\circ}$C. From Ref.~\cite{Garutti:2017ipx}.}
\label{Fig:IV}
\end{figure}

Effects of radiation damage in silicon have been introduced in Sec.~\ref{sec:sirad}. 
The main macroscopic effect in SiPMs reported in several studies is the significant increase of dark current ($I_{dark}$) below and above breakdown. For SiPM operation this translates in an increase of dark count rate (DCR). The  scaling of $I_{dark}$ and DCR, with particle type, energy and irradiation fluence is not yet well documented. Some examples from literature are reported in the following. Fig.~\ref{Fig:IV} shows an example of current-voltage curves for a SiPM irradiated with neutrons, and operated at -30$^{\circ}$C. In the region of unit gain (V~$\sim$ 5~V) the dark current increases by about three orders of magnitude after $\Phi_{eq} = 5 \cdot 10^{14}$~cm$^{-2}$, whereas above breakdown voltage the increase is more than six orders of magnitude. For too high over-voltages the measurements at the highest fluences are affected by saturation due to the current limit of the power supply.      

\begin{figure}[h]
\centering
\includegraphics[width=0.65\linewidth]{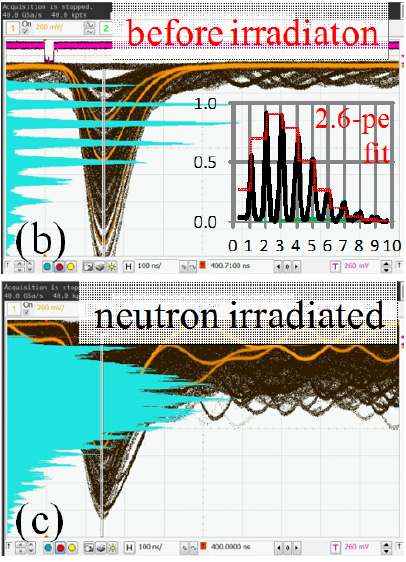}
\caption{Single photoelectron charge signal pulses at  $V-V_{bd}$=3~V (b) before irradiation, (c) after neutron irradiation to $\Phi_{eq} = 10^{9}$~cm$^{-2}$. From Tsang Ref.~\cite{Tsang2016}.}
\label{Fig:SPE}
\end{figure}

Another major effect of radiation on SiPM is the loss of single photon counting resolution due to the noise increase. This effect is reported in all studies dealing with bulk damage in SiPMs, and occurs at different fluences depending on the SiPM design, operating temperature, type of irradiation, and other parameters. It is safe to state that between     
$\Phi_{eq} = 10^{9}-10^{10}$~cm$^{-2}$, all SiPMs lose single photon counting resolution when operated at room temperature. Fig.~\ref{Fig:SPE} shows an example of a pulse-height spectrum for a SiPM before irradiation in which the photo-electron peaks are resolved. In the  spectrum after $\Phi_{eq} = 10^{9}$~cm$^{-2}$, the photo-electron peaks are no longer resolved at room temperature. Tsang and co-authors demonstrate in Ref.~\cite{Tsang2016} that the same SiPM cooled at 84~K still resolves single photons. 

\begin{figure}[h]
\centering
\includegraphics[width=0.9\linewidth]{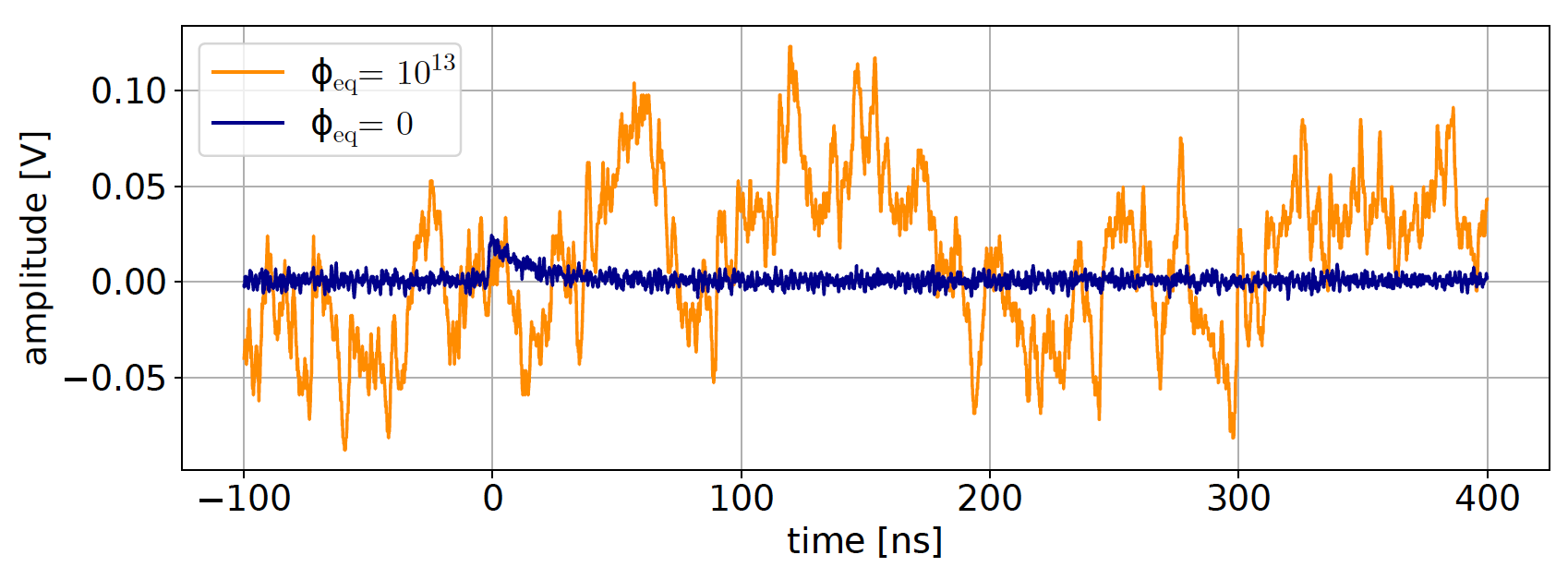}
\includegraphics[width=0.9\linewidth]{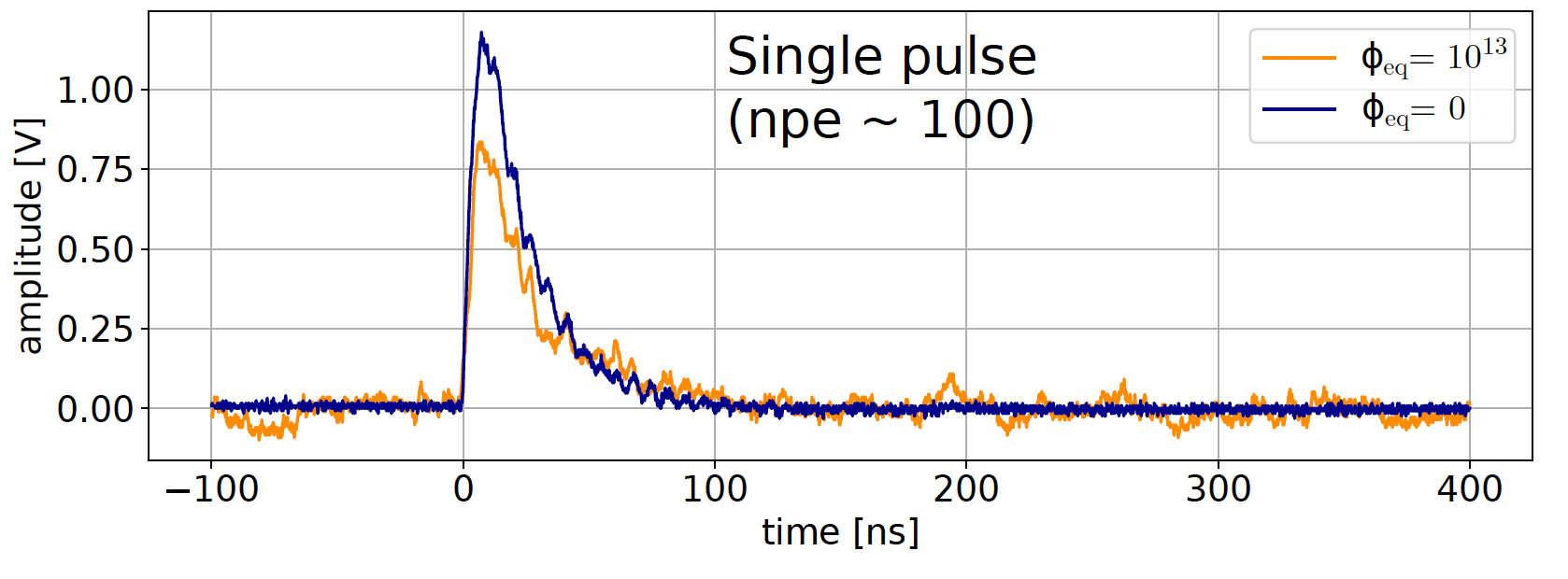}
\caption{Waveforms of a KETEK SiPM operated at -30$^{\circ}$C and 2.5 V overvoltage.  The blue lines are before and the orange lines after irradiation with neutrons to $\Phi_{eq} = 10^{13}$~cm$^{-2}$. The waveforms in the top plot are taken in the dark, those in the bottom plot are taken with the SiPM under illumination with an LED light intensity of approximately 100 photoelectrons. From S. Cerioli doctoral thesis in progress.}
\label{Fig:waveforms}
\end{figure}

Another example is given in Fig.~\ref{Fig:waveforms}. The top plot compares waveforms acquired in the dark with a KETEK SiPM operated at -30$^{\circ}$C. The single photo-electron peak visible before irradiation (blue curve) cannot be resolved in the noisy baseline after neutron irradiation to $\Phi_{eq} = 10^{13}$~cm$^{-2}$. The bottom plot demonstrates that the device is still a functional photo-detector for larger light intensity. 

For the characterization of SiPMs this implies that when the single photon resolution is lost, new methods need to be applied to extract the SiPM characteristic parameters, e.g. gain, DCR, correlated noise, PDE, etc...\footnote{For a detailed overview of standard SiPM characterization methods we refer to Ref.~\cite{Klanner:thisVolume} within this review volume.}.   
A straight forward method allowing to investigate possible changes in the product of the factors contributing to the SiPM response is to plot the normalized ratio of photo-currents\footnote{The photo-current is the current in the presence of light subtracted from the dark current: $I_{photo}=I_{light}-I_{dark}$, and it is given by $I_{photo}= q_0 N_{\gamma}~G ~ECF ~PDE $.} before ($\Phi =0$) and after ($\Phi$) irradiation~\cite{Garutti2016:IEEE}.

The ratio of normalized photo-currents is defined as:
\[
R=\frac{I_{photo}^{norm}(\Phi)}{I_{photo}^{norm}(\Phi =0)} = \frac{( G_{\Phi} ~ECF_{\Phi}~ PDE_{\Phi}) / (M_{\Phi}~ QE_{\Phi})}{\underbrace{(G_0 ~ECF_0 ~PDE_0 )}_\text{$V>V_{bd}$} / \underbrace{(M_0 ~QE_0)}_\text{$V \ll V_{bd}$}},
\]
with the Excess Charge Factor (ECF) defined as the ratio between the mean values of measured $\langle  N_{pe} \rangle$ and primary produced Geiger avalanches, $\langle N_{pG} \rangle$:  
\begin{equation}
ECF=\frac{\langle N_{pe} \rangle}{\langle N_{pG} \rangle}.
\label{eq:ecf}
\end{equation}
where it is assumed that the number of primary produced avalanches follows the same distribution as the impinging light, and that the light source emission follows a Poisson distribution.

The photo-current is normalized to its value, $I_{photo} = q_0 ~N_{\gamma}~M ~QE $, at $V<<V_{bd}$ where the amplification gain can be considered $M \approx 1$. A value of $R=1$ for $V\ll V_{bd}$ confirms that the product of amplification factor and quantum efficiency ($M~QE$) of the SiPM is unchanged after irradiation. For $V>V_{bd}$ a value $R=1$ indicates that the product of gain,  correlated noise, and PDE is not changed after irradiation. Any deviation from $R=1$ indicates that one or more of these parameters are affected (an example of such result is discussed in Sec.~\ref{sec:had}). 

\begin{figure}[h]
\centering
\includegraphics[width=0.90\linewidth]{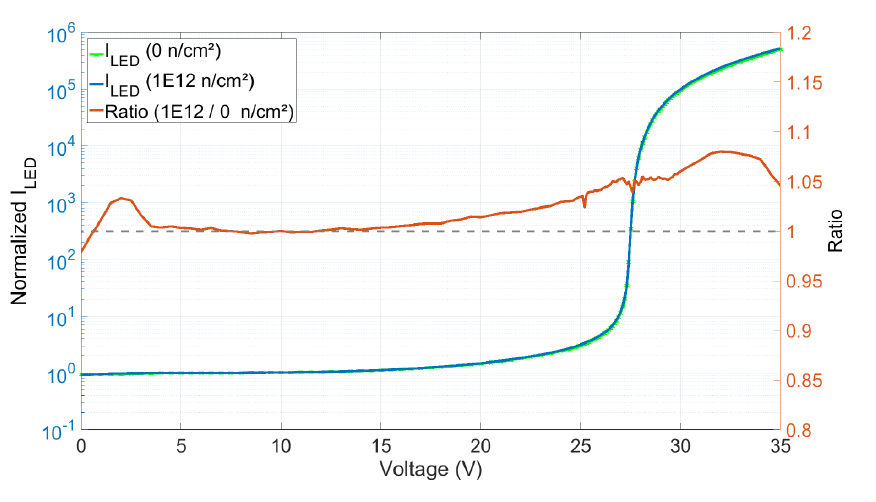}
\caption{The normalized LED photo-current of a non-irradiated SiPM and a SiPM irradiated to $\Phi_{eq} = 10^{12}$~cm$^{-2}$, taken from Ref.~\cite{Garutti2016:IEEE}. Although the normalized photo-current spans five orders of magnitude, the ratio of the measurements before and after irradiation shows a variation of less than 7\%. }
\label{Fig:photocurrent}
\end{figure}

Fig.~\ref{Fig:photocurrent} shows $R$ for a KETEK SiPM irradiated with neutrons to $\Phi_{eq} = 10^{12}$~cm$^{-2}$, and  operated at -30 $^{\circ}$C. $R$ agrees well with unity below breakdown and above breakdown is at most 7\% higher than unity. This suggests that after irradiation the product $G_{\Phi}~ECF_{\Phi}~PDE_{\Phi}$ has not changed by more than 7\% with respect to the original value. 
This and other observations discussed in Sec.~\ref{sec:had} suggested to split the discussion of bulk damaged SiPM in two fluence regions: 
Medium-fluence irradiations, for which the values of $G$, $ECF$, $PDE$ can be assumed not to change, and high-fluence irradiation for which the changes in these SiPM parameters must be taken into consideration in the interpretation of results. 

In the following discussion we present studies of radiation damage on SiPMs separated in:
Surface radiation damage, induced by X-ray irradiations (Sec.~\ref{sec:Xray}), and bulk radiation damage, induced by high-energetic electrons and gammas (Sec.~\ref{sec:em}), or by protons and neutrons (Sec.~\ref{sec:had}). It should be noted that aside of the latter all other particle types induce both surface and bulk damage in SiPMs, but the separation of these effects is not further discussed. 

\subsection{Surface radiation damage - X-ray photons}
\label{sec:Xray}
Electrons and photon with energies below the threshold for 
bulk defects ($\sim$300~keV) generate only defects in the dielectrics, at the Si-SiO$_2$ interface and at the interface between dielectrics. About $\sim$18~eV are required to generate an e-h pair in SiO$_2$. This is the typical case for X-ray irradiation where photon energies are usually in the range $<$100~keV. Details of the expected surface damage effects are discussed in Sec.~\ref{sec:sirad-iel}. 

The SiPM electric parameters ($R_q, C_{pix}, \tau$) are extracted from measurements of current-voltage, and  capacitance/conductance-voltage for frequencies between 100~Hz and 2~MHz just below $V_{bd}$. 
Breakdown voltage, gain, dark-count rate, cross-talk probability and pulse shape above the breakdown voltage are also investigated. The values of pixel capacitance and quenching resistor do not alter after irradiation. Their product $\tau_{RC} = C_{pix}R_q$ defines the pixel recovery time. This quantity is compared to $\tau_{exp}$ extracted as exponential slope of the signal from current transients. For this device the two definitions of $\tau$ are found to agree within experimental errors. 

After 20~MGy the current below $V_{bd}$ increases by three orders of magnitude compared to the value before irradiation. This effect is ascribed to an increase of the surface-generation current from the Si-SiO$_2$
interface. The same increase is also visible in the current above breakdown, whereas the dark-count rate increases only by an order of magnitude at the same dose. The difference between dark-current and dark-count rate increase indicates that the large fraction of the current is not amplified in the multiplication region. 
These studies indicate that when developing SiPMs for applications in high dose X-ray environment, one must pay attention to the surface design, to minimize the increase of surface current due to radiation and to prevent that the surface current reaches the multiplication region. 

\subsection{Bulk radiation damage - electrons and gammas}
\label{sec:em}
SiPMs exposed to $^{60}$Co gamma irradiation ($E_{\gamma}$~=~1.33, 1.17 MeV) will experience a combination of surface and bulk defects, and are expected to have similar effects as SiPMs irradiated with electrons or positrons. 

Matsubara and co-authors in Ref.~\cite{Matsubara:2006zz} have irradiated a prototype SiPM from Hamamatsu (Type No. T2K-11-100C) under bias up to 240~Gy of $^{60}$Co $\gamma$-rays and measured the dark current, dark-count rate, gain, and cross talk. Whereas gain and cross talk did not significantly change with dose, large dark-count pulses and localized spots with leakage current along the outer edge of the active region and the bias lines were observed for about half an hour after irradiation for doses above 200~Gy. Renker commented in Ref.~\cite{RenkerP04004} that very likely this effect was caused by accumulated and stationary charges at the Si-SiO$_2$ interface generated by the breakup of SiO$_2$ molecules.
The authors also observed that immediately after powering the SiPM had large dark counts with an amplitude corresponding to a signal of more than 10 photo-electrons. After a couple of minutes, all signals with large amplitudes and most of the dark current disappeared. This phenomenon reappears after power cycling the device to zero and back to operating voltage.

\begin{figure}[h]
\centering
\includegraphics[width=0.9\linewidth]{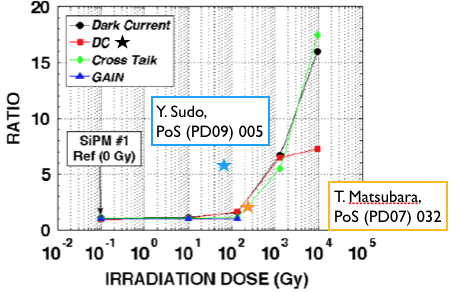}
\caption{Ratio of measured quantities vs. irradiation dose. Modified from~\cite{Pagano:2014bua}. The red squared indicate the Dark Counts (DC) from  ~\cite{Pagano:2014bua} taken at $\Delta V$~=~3~V ($\Delta V/V_{bd}$ = 11\%). The two additional star points indicate measurements from~\cite{Matsubara:2006zz} (orange) and ~\cite{Sudo2009} (blue).}
\label{Fig:Pagano}
\end{figure}

The study was extended to higher doses by Lombardo et al. in Ref.~\cite{Pagano:2014bua}, where SiPMs from ST Microelectronics (400 pixels, 0.64~mm$^2$ active area, 0.47 fill factor) were irradiated with a $^{60}$Co source to 0, 10, 136~Gy, and 1.3, 9.4~kGy. Current-voltage and transient characteristics were recorded. Single photo-electron peaks are visible in the integrated spectrum up to 136~Gy. Above 1~kGy the noise increase is such that no structure can be distinguished in the pulse-height spectra, so the gain of the device can no longer be measured. The increase of dark current, dark count rate and cross talk as function of the dose is shown in Fig.~\ref{Fig:Pagano}.  The authors explain that the method used to determine cross talk, using the fraction $CT(\%) \sim \frac{DC_{1.5}}{DC_{0.5}}\times 100$  may not be suitable for doses larger than 1.3~kGy.
The measurement of dark count rate at 9.4~kGy may also be affected by  saturation due to the method used, namely to count pulses above a certain threshold in a one second long transient. For a recovery time $\tau = R_qC_{pix} \sim$~100~ns, the counting method is supposed to saturate at about 10$^7$ counts/sec, and this is the order of magnitude reached after 9.4~kGy. 
The increase of dark counts for the highest dose measurement of Matsubara et al. agrees qualitatively with the trend presented in Ref.~\cite{Pagano:2014bua} (orange star in Fig.~\ref{Fig:Pagano}). 

In Ref.~\cite{RenkerP04004} it is reported, that several SiPMs have been irradiated up to 500~Gy by a $^{60}$Co-source without applying a bias voltage during irradiation. No evidence for large pulses has been found after the irradiation.  In Ref.\cite{QIANG2013234}, in which the radiation hardness of Hamamatsu SiPMs was investigated, footnote 1 states: "An early irradiation test on SiPMs using a series of high activity $^{137}$Cs-sources in Jefferson Lab showed that SiPMs are insensitive to electromagnetic radiation and there was no significant change in performance of SiPMs up to 2~krad of gamma irradiation.".

\begin{figure}[h]
\centering
\includegraphics[width=0.7\linewidth]{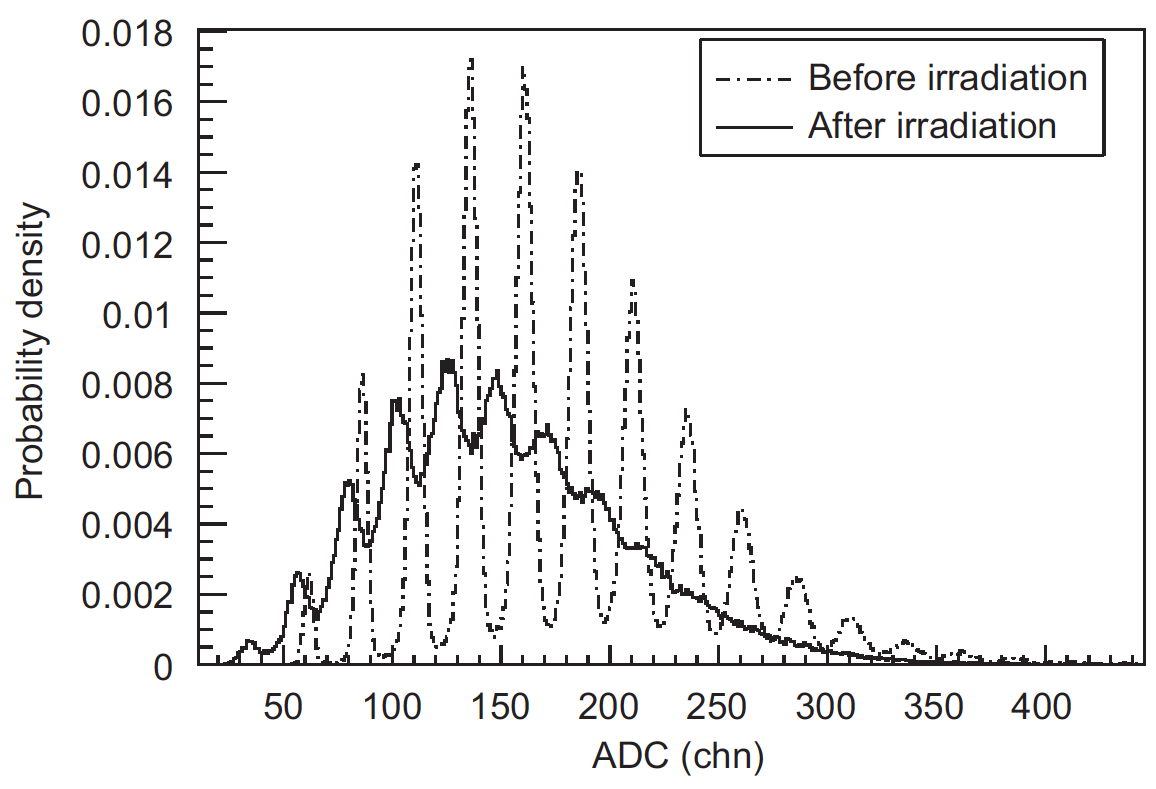}
\caption{Single photoelectron spectra for low light intensity illumination of a Photonique green-sensitive SiPM before and after irradiation with 14~MeV electrons, to $\Phi = 3.1 \cdot 10^{11}$~cm$^{-2}$. From Achenbach Ref.~\cite{SANCHEZMAJOS2009506}.}
\label{Fig:SPEelec}
\end{figure}

Achenbach et al. have irradiated green-sensitive SiPMs
(SSPM-0701BG-TO18 from Photonique S~A, Geneva) with 14~MeV electrons to 
fluences between 3.1$\cdot$10$^{11}$ cm$^{-2}$ and 3.8$\cdot$10$^{12}$ cm$^{-2}$ and observed a large increase in dark-count rate and a decrease in effective gain, see Ref.~\cite{SANCHEZMAJOS2009506}.
After irradiation the single photo-electron peaks are still visible as demonstrated in Fig.~\ref{Fig:SPEelec}. Surface effects were deemed responsible for the observed  shift in the pedestal position, as consequence of the increase in leakage current.
The SiPM response to a medium-intensity laser light was investigated. A progressive reduction of signal charge is observed as a function of the irradiation dose. The paper does not attempt to disentangle the effects of decrease in gain, PDE in the single pixel or increase of pixel occupancy.
A similar study was conducted in Ref.~\cite{MUSIENKO2007433}, using 28~MeV positrons with fluences up to 8$\cdot$10$^{10}$~cm$^{-2}$. PDE, gain, dark current and dark count rate were measured at room temperature as a function of bias voltage before and 2 days after irradiation. No change of the PDE and
gain were found. As in the other publication the dark current and counts are shown to increase as function of dose. The authors introduce a new variable to quantify the effect of radiation damage on the dark count rate ($DCR$):\\
\begin{equation}
\Delta DCR_{norm} =  \frac{DCR(\Phi)-DCR(\Phi=0)}{ A\cdot PDE_{515nm}}
\label{Eq:DN}
\end{equation}
where $A$ is the active area of the device, and the PDE is calculated at 515~nm. 
\begin{figure}[h]
\centering
\includegraphics[width=0.9\linewidth]{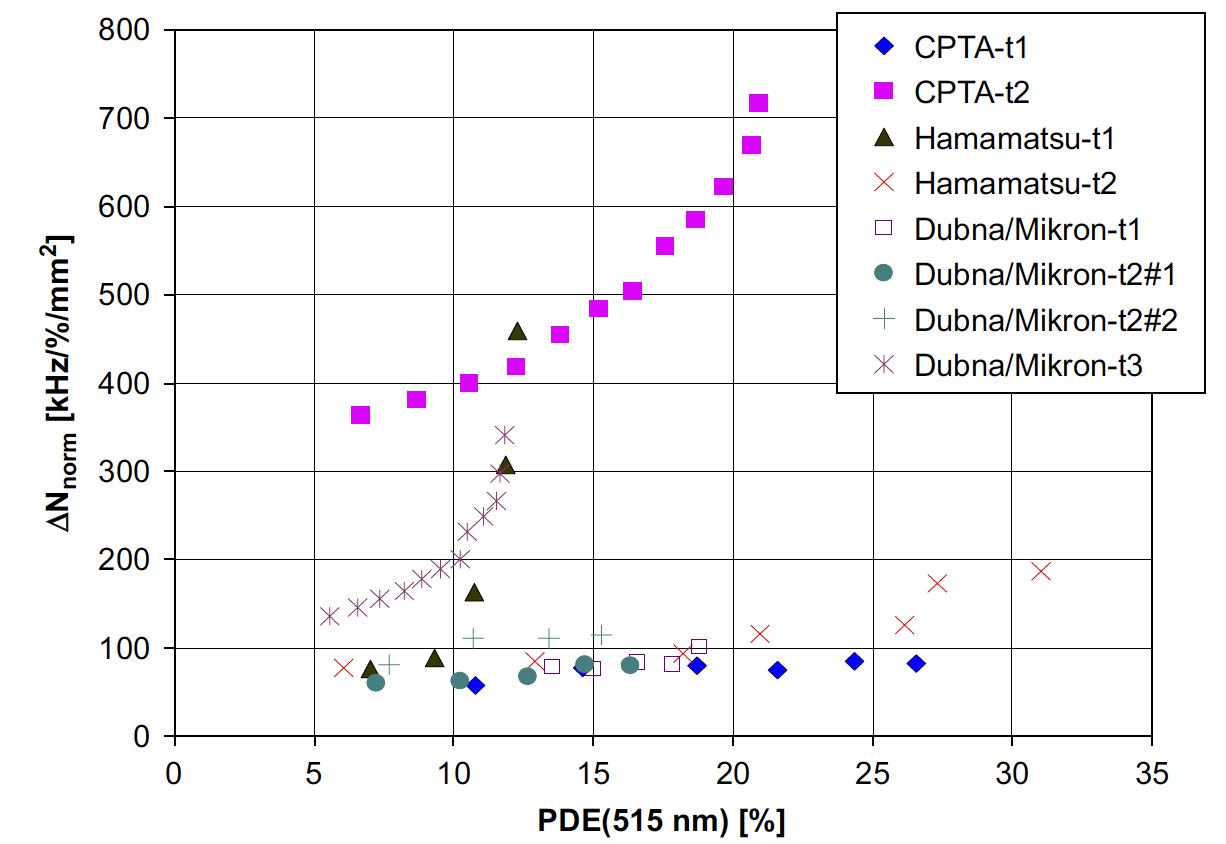}
\caption{Increase of dark count rate after irradiation, normalized to active area and PDE as expressed in Eq.~\ref{Eq:DN} and plotted vs the PDE of several SiPM irradiated with 28 MeV positrons to $\Phi = 8\cdot 10^{10}$~cm$^{-2}$. The variable labeled $\Delta N_{norm}$ on the y-axis is the asme as defined in Eq.~\ref{Eq:DN}. Plot taken from Ref.~\cite{MUSIENKO2007433}.}
\label{Fig:positron}
\end{figure}
By plotting $\Delta DCR_{norm}$ vs PDE, the authors argue that one can recognize devices with similar depletion layer thickness, and that good quality SiPM should have small PDE dependence of $\Delta DCR_{norm}$ (labeled $\Delta N_{norm}$ in Fig.~\ref{Fig:positron}). Unfortunately, due to lack of information it is not possible to extract the same quantity from other electromagnetic irradiation results so that the comparison between positron, electron and gamma radiation damage effects remains purely qualitative. 






\subsection{Bulk radiation damage - hadrons}
\label{sec:had}
SiPMs have been irradiated by several groups with neutrons~\cite{Nakamura09,QIANG2013234,Andreotti:2013nra,Musienko:2015lia,Heering:2016lmu,Garutti:2017ipx,CentisVignali:2017zpz,Musienko:2017znn,Cattaneo17} and protons~\cite{Matsumura:2009he,DANILOV2009183,BOHN2009722,LI201663,MUSIENKO200987,SANCHEZMAJOS2009506}, and their properties investigated. In this review we combine results on hadron irradiation without distinguishing particle type, since no dedicated studies have been performed so far on SiPMs on this topic. Fluences are quoted in 1 MeV neutron equivalent, according to the NIEL scaling explained in Sec.~\ref{sec:sirad-niel}.   
We group the studies in two categories: Medium-fluence irradiations, up to $\Phi_{eq} \approx 10^{12}$~cm$^{-2}$, and high-fluence irradiation to higher fluences. For medium fluences the majority of the studies report no significant change in SiPM parameters $V_{bd}$, $R_q$, $C_{pix}$, PDE, gain. Whereas, dark current (DC) and dark count rate (DCR) significantly increase proportionally to the fluence. For high fluences, $\Phi_{eq} > 10^{12}$~cm$^{-2}$, also other parameters are affected; but most importantly, the methods to determine the parameters are strongly affected by the high dark current and their limitations need to be discussed carefully.

\subsubsection{Effects on $R_q$, $C_{pix}$, $V_{bd}$}
\label{sec:rq}
\begin{figure}[h]
\centering
\includegraphics[width=0.9\linewidth]{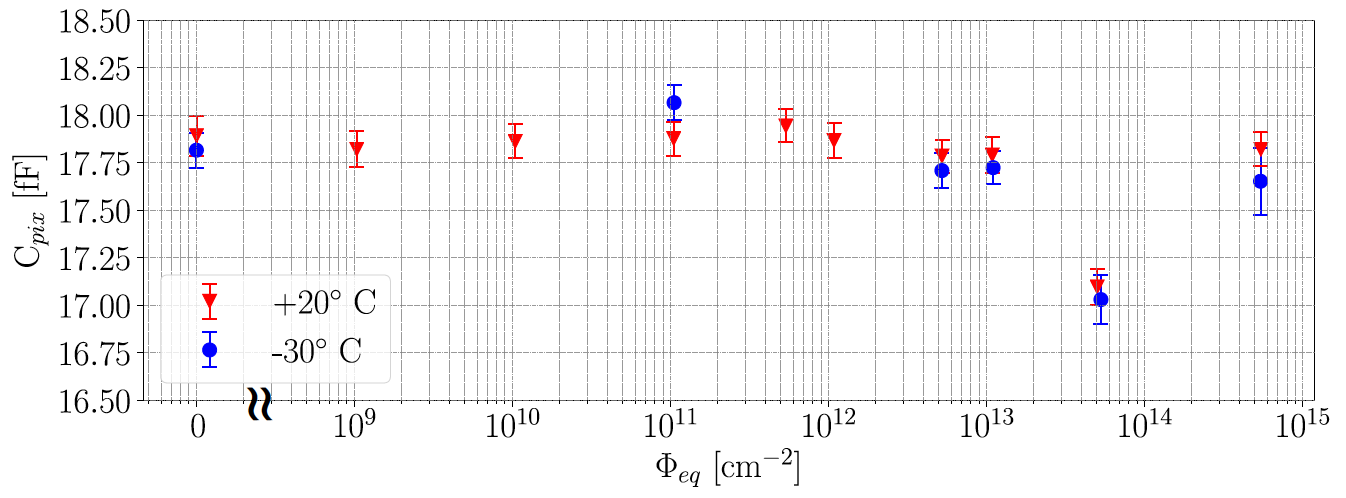}
\includegraphics[width=0.9\linewidth]{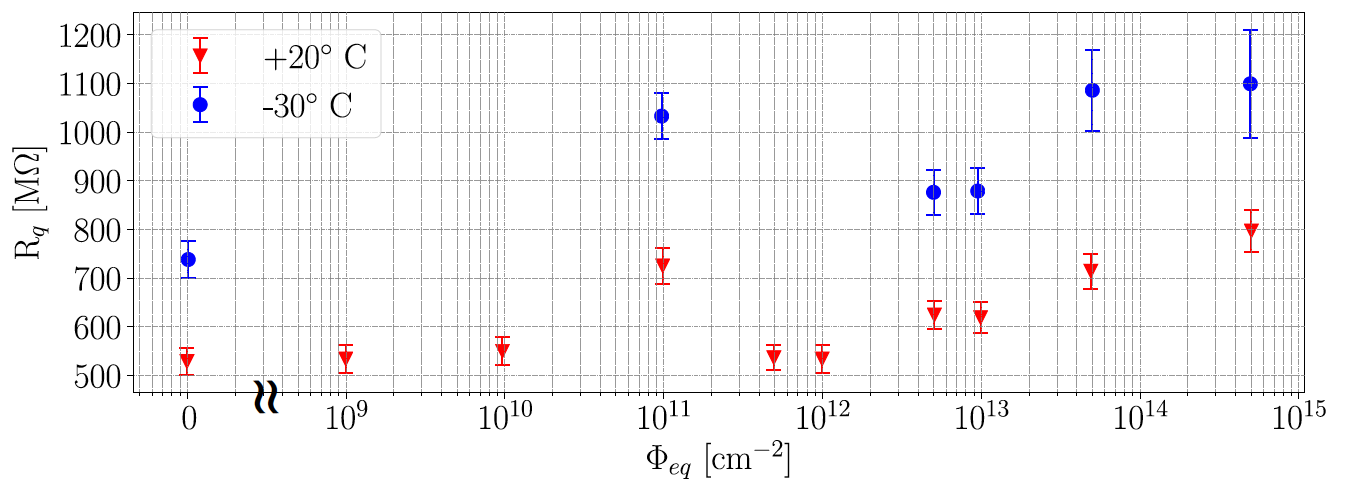}
\caption{SiPM electrical parameters of a KETEK SiPM (15 $\mu$m pixel size) as the function of neutron fluence measured at +20$^{\circ}$C and -30$^{\circ}$C. (Top) Pixel capacitance, $C_{pix}$, and (bottom) quenching resistance, $R_q$.}
\label{Fig:cpix}
\end{figure}

The values of $R_q$, $C_{pix}$ can be extracted from impedance-frequency scans below breakdown, both before and after irradiation. 
In Fig.~\ref{Fig:cpix} results are presented exemplary for KETEK SiPMs irradiated with neutrons. These plots are an extension to higher fluences of the analysis already published in Ref.~\cite{Garutti2016:IEEE}. From the C-V measurements below the breakdown voltage, which were taken at 25 frequencies between 100 Hz and 2 MHz, the SiPM electrical parameters have been determined using a simple R-C model. It is found that the value of $C_{pix}$ neither depends on temperature nor on neutron fluence, whereas the value of $R_q$ increases for $\Phi_{eq} > 10^{12}$~cm$^{-2}$. As expected for a poly-Si resistor, $R_q$ increases with increasing temperature. 
Ref.~\cite{MUSIENKO200987} shows similar results for proton irradiation up to $\Phi_{eq} = 2\cdot 10^{10}$~cm$^{-2}$.

In the absence of single-photon counting capability the turn-off voltage cannot be extracted from the linear relation between the excess bias voltage and the gain, $G=C_{pix}(V_{bias}-V_{bd})/q_0$\footnote{More correctly this expression should read $G=(C_{pix}+C_q)(V_{bias}-V_{off})/q_0$, taking into consideration the difference between the turn-on voltage for the Geiger avalanche (the breakdown voltage), and the turn-off voltage demonstrated in Ref.~\cite{Chmill:2016msk}.}. I-V measurements are use to determine $V_{bd}$ in this case. It is recommended to use I-V measurements taken under illumination of the SiPM with a source of light, either pulsed or continuous, and to apply a simple quadratic interpolation to the minimum of the inverse logarithmic derivative of the current, or analogously to the maximum of the logarithmic derivative. This procedure is stable against observed changes in the I-V dependence of dark noise below breakdown before and after irradiation possibly due to surface current, and also against the large uncertainty in the current measurement, which may affect I-V below breakdown at low temperatures or for low fluences.  

One should keep in mind that the observable obtained with the I-V method, is not the same as that obtained from gain vs voltage linear regression~\cite{Chmill:2016msk}. In the first case the turn-on voltage for the Geiger avalanche is measured (the breakdown voltage), whereas the former gives access to the turn-off voltage, which is the voltage relevant for the gain calculation in SiPMs. The two values can be significantly different, and it is not known if they both react to radiation damage in the same way. 

\begin{figure}[h]
\centering
\includegraphics[width=0.9\linewidth]{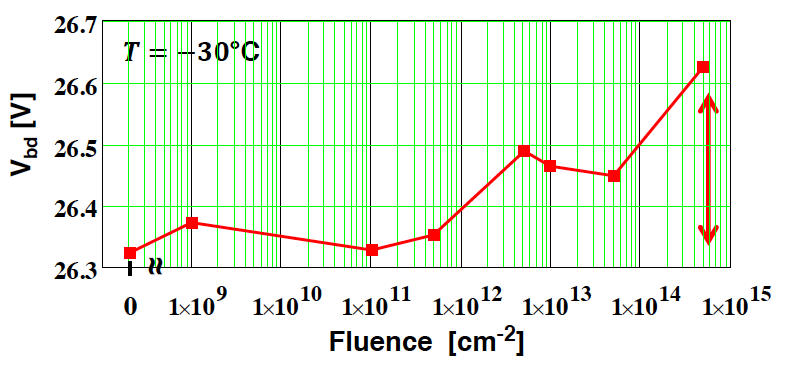}
\caption{$V_{bd}$ of a KETEK SiPM (15 $\mu$m pixel size) as the function of neutron fluence measured at -30$^{\circ}$C.}
\label{Fig:vbd}
\end{figure}

Ref.~\cite{Heering:2016lmu} presents the first evidence of a $V_{bd}$ shift due to radiation damage. Hamamatsu SiPMs exposed to high-fluence proton irradiation, $\Phi_{eq} = 6\cdot 10^{12} - 2\cdot 10^{14}$~cm$^{-2}$, show a linear increase of $V_{bd}$ with fluence, up to 4~V.  
Fig.~\ref{Fig:vbd} presented in Ref.~\cite{Klanner:ICASIPM} shows $V_{bd}$ constant within uncertainties up to $\Phi_{eq} = 5\cdot 10^{13}$~cm$^{-2}$  and a possible increase of 300~mV after $\Phi_{eq} = 5\cdot 10^{14}$~cm$^{-2}$ for KETEK SiPMs. 
The difference in the magnitude of the $V_{bd}$ shift could be explained by the difference in the multiplication layer width. For Hamamatsu SiPM the multiplication layer is wider ($d\sim$~2~$\mu m$) than for KETEK SiPMs ($d\sim$~1.2~$\mu m$). It appears that thinner multiplication regions are less sensitive to breakdown voltage shifts by radiation damage. This could be linked to electric charges creation inside the depleted volume. If one assumes uniform acceptor generation during irradiation, the $V_{bd}$ shift should be proportional to the square of the depletion depth, as proposed in Ref.~\cite{Musienko:2000im} for breakdown voltage shifts in APDs. 


\subsubsection{Effects on $I_{dark}$ and DCR}
\label{sec:DCR}
\begin{figure}[h]
\centering
\includegraphics[width=0.9\linewidth]{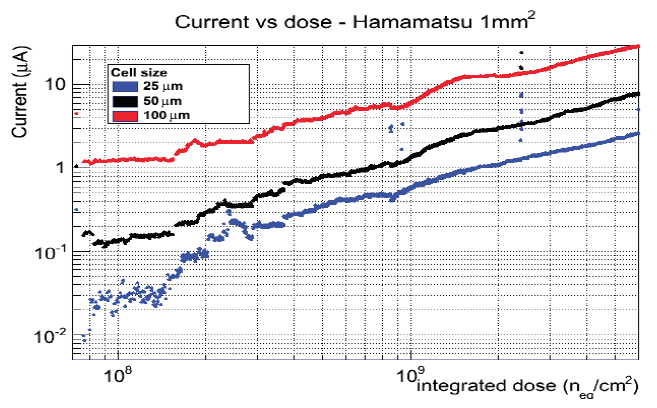}
\caption{Dark current vs neutron dose for 1~mm$^2$ HAMAMATSU SiPMs with different pixel size. The blue line is
for the 25~$\mu$m pixel, black 50~$\mu$m and red 100~$\mu$m. From Ref.~\cite{Andreotti:2013nra}. In the paper similar measurements for AdvanSiD and SensL are also reported.}
\label{Fig:pixsize}
\end{figure}

Up to $\Phi_{eq} \approx 10^{10}$~cm$^{-2}$ the typical increase of dark current is by a factor 30-100, confirmed by various studies on different devices.   
Cibinetto et al.~\cite{Andreotti:2013nra} have irradiated samples from various producers with neutron fluences up to $\Phi_{eq} = 6.2\cdot 10^{9}$~cm$^{-2}$, and show that the relative increase in current is independent of the pixel size despite the significant higher current values for larger pixel size or capacitance (see Fig.~\ref{Fig:pixsize}). 

Quiang et al.~\cite{QIANG2013234} present a linear correlation between dark current and dark count rate, expected if one assumes that all the dark current is generated by carriers that traverse the multiplication region initiating a Geiger discharge: 
\begin{equation}
I_{dark} = q_0 ~G~ECF~DCR.
\label{eq:idark}
\end{equation}
Note that in the absence of correlated noise (cross-talk and afterpulse)  $ECF=1$. For SiPMs typically $ECF<1.2$, with values increasing as a function of the excess bias voltage. 
Fig.~\ref{Fig:photocurrent} demonstrates that the assumption that $ECF$ is independent of fluence is justified for medium-fluences.  
For high-dose irradiation the SiPM current can easily exceed hundreds of micro-ampere\footnote{Note that as a consequence of the significant current increase the bias voltage applied to the SiPM, $V_{bias}$, must be calculated correcting the power supply voltage $V_{PS}$ by the voltage drop over the series resistance of the biasing circuit, $V_{bias}=V_{PS}-I_{leak}\cdot R$. Where $R$ is a value specific for a given set-up. This can cause voltage drops of several volts depending on the circuit for current values in the range of mA.} also when cooling the device below zero degree. These high currents can cause a local warm up of the pixel. The pixel temperature as function of $I_{dark}$ is hard to estimate, but a current-induced temperature increase could possibly be responsible for a change in $V_{bd}$ as function of fluence. 
 
For non-irradiated SiPMs the DCR is often determined from the assumption of Poisson distributed dark count signals in a time interval $t_{gate}$ (e.g. the width of the gate), as:
\[
DCR=-\frac{\ln(P_0)}{t_{gate}}
\]
with $P_0$ the fraction of events with pulse areas corresponding to $N_{pe} < 0.5$. 
In irradiated SiPM this method can no longer be applied due to the loss of single photon resolution. Other methods have been tested and should be investigated more thoroughly. 
In Ref.~\cite{Garutti:2017ipx} the DCR of irradiated SiPM is scaled from the DCR before irradiation using the ratio of the dark current:
\begin{equation}
DCR_{IV}(\Phi) = \frac{I_{dark}(\Phi)}{I_{dark}(\Phi = 0)}~ DCR(\Phi = 0).
\label{eq:DCR1}
\end{equation}
However, this relation holds only under the assumption that gain and ECF are unchanged after irradiation, which has been confirmed in Fig.~\ref{Fig:photocurrent} for $\Phi_{eq} \leq 10^{12}$~cm$^{-2}$.

Another method is to define the DCR from the rms of the charge spectrum obtained with no illumination, $\sigma_0$. An often used definition is:
\[
DCR^* \propto \frac{\sigma^2_{0}}{\Delta t}.
\]
If $\sigma_{0}$ has the units of charge it should be divided by $q_0 ~G$ to obtain the DCR in units of counts/second. This approximation does not take into account the non-Poisson distribution of the SiPM signal. The time interval $\Delta t$ is sometimes taken to be the integration gate, with no further assumption on its relation to the signal shape, which may not be correct in case of short gates. Following the derivation presented in Ref.~\cite{Klanner:thisVolume} the relation between DCR and $\sigma^2_{0}$ for the simple case of a SiPM signal with one recovery time component $\tau = R_q C_{pix}$ is: 
\begin{equation}
DCR_{rms} = \frac{\sigma^2_{0}}{q_0^2~G^2 ~ENF~ECF^2 \cdot \left(t_{gate}+\tau \left( e^{-\frac{t_{gate}}{\tau}} -1 \right)\right)}.
\label{eq:DCR2}
\end{equation}
where the expression in brackets comes from the variance of the distribution of $N_{DC}$ dark photo-electron pulses in a time interval $\Delta t_1$, defined such that $N_{DC} = DCR \cdot \Delta t_1 $. In the case of a signal much shorter than the integration gate,  $\tau << t_{gate}$, the expression in brackets simplifies to $t_{gate}$. 
For more complex signal shapes another expression needs to be derived.

The Excess Charge Factor (ECF) is defined in Eq.~\ref{eq:ecf} and the Excess Noise Factor (ENF) is defined as:
\begin{equation}
ENF=\left(\frac{\sigma_Q}{\langle Q \rangle}\right)^2 \left( \frac{\langle Q_P \rangle }{(\sigma_Q)_P} \right)^2 
   = \left(\frac{\sigma_Q}{\langle Q \rangle} \right)^2 \langle N_{pG} \rangle,
\label{eq:enf}
\end{equation}

\begin{figure}[h]
\centering
\includegraphics[width=0.90\linewidth]{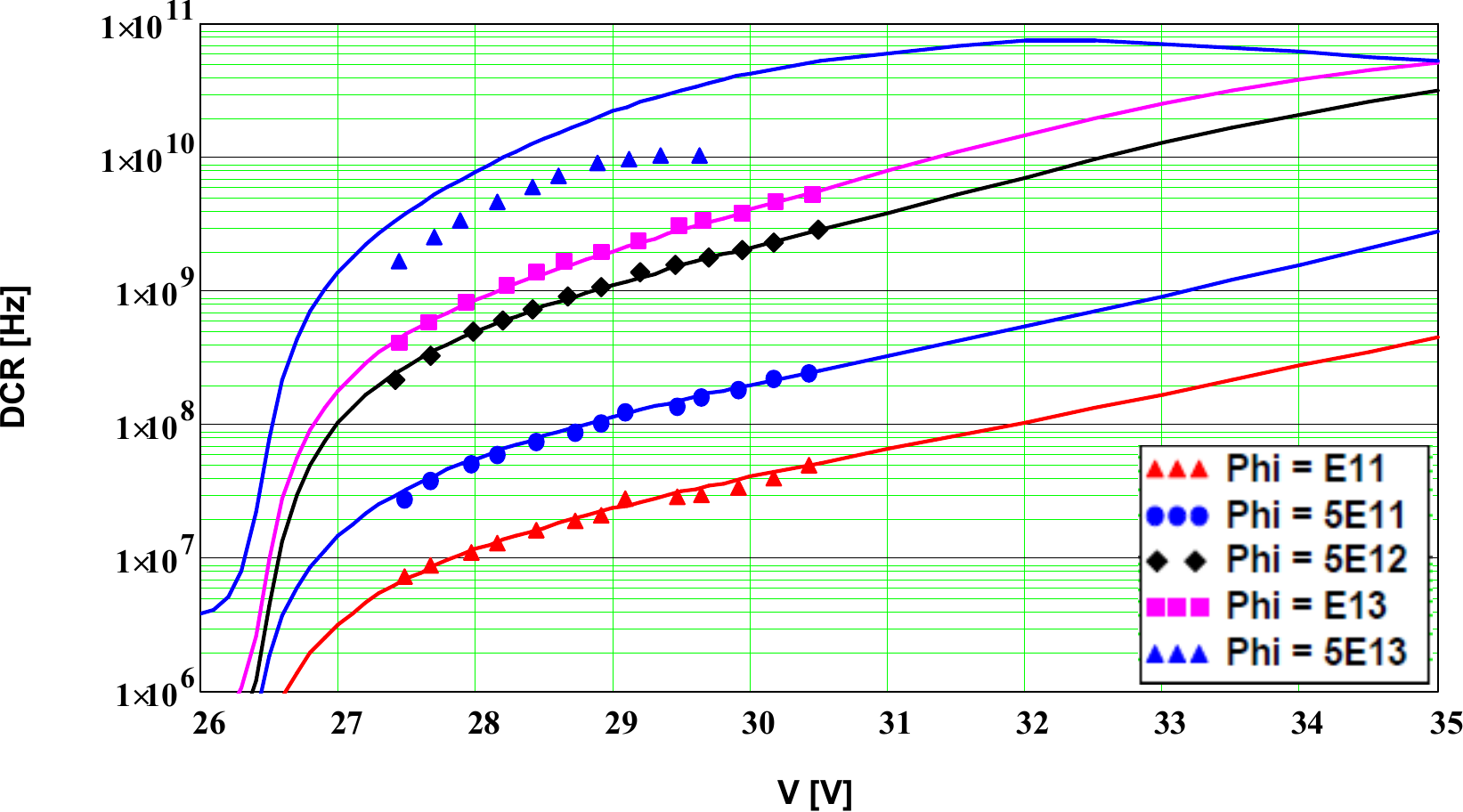}
\caption{Comparison of DCR as extracted from Eq.~\ref{eq:idark} (lines) and Eq.~\ref{eq:DCR2} (points) for KETEK SiPMs irradiated with neutrons and operated at -30$^{\circ}$C. No annealing is performed. Adapted from \cite{Klanner:ICASIPM}. }
\label{Fig:DCR}
\end{figure}

To extract DCR from Eq.~\ref{eq:DCR2} for an irradiated SiPM requires the assumption that gain and correlated noise are not affected by radiation damage. In addition, it is assumed that the signal shape does not change significantly after irradiation, otherwise different gate lengths may be needed for measurements before and after irradiation. 
Fig.~\ref{Fig:DCR} shows a very preliminary study comparing DCR values obtained using Eq.~\ref{eq:idark} and Eq.~\ref{eq:DCR2} on the same SiPM. 
For fluences between $10^{11}$~cm$^{-2}$ and $10^{13}$~cm$^{-2}$ the agreement is reasonably good. For the highest fluence of $\Phi_{eq} = 5  \cdot 10^{13}$~cm$^{-2}$ a saturation effect is observed in the transient RMS method, due to high pixel occupancy. This effect is not implemented in the analysis yet. 
The DCR data obtained from transient measurements are affected by about 30\% correlated systematic uncertainties so that at this point it is only possible to say that the shapes are well in agreement, but the absolute values are subject to further investigations. Several of the assumptions still have to be verified, an example is the possible effect of an increase in after-pulsing and late cross-talk, which may mean that the ENF and ECF are different between a 100~ns gate and the current measurement (infinite gate length).

\subsubsection{Effects on signal, PDE, gain}
\label{sec:gain}

\begin{figure}[h]
\centering
\includegraphics[width=0.9\linewidth]{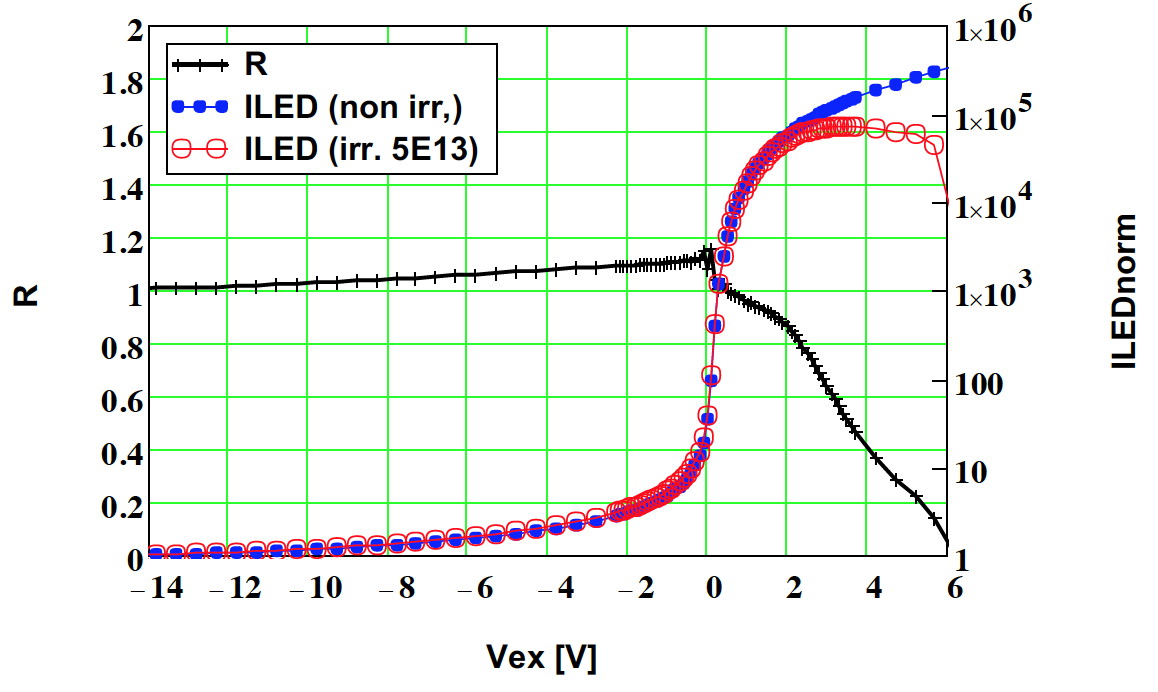}
\caption{The normalized LED photo-current of a non-irradiated SiPM and a SiPM irradiated to $\Phi_{eq} = 5\cdot 10^{13}$~cm$^{-2}$. The KETEK SiPM is operated at -30$^{\circ}$C. From Ref.~\cite{CentisVignali:2017zpz}}
\label{Fig:photocurrent2}
\end{figure}

As already argued, generally in irradiated devices the PDE and gain cannot be directly measured due to the noise increase. 
With the analysis of the normalized photo-current as presented in  Fig.~\ref{Fig:photocurrent}, one can determine the product $PDE(\Phi) ~G(\Phi) ~ECF(\Phi)$. As long as $R=I_{photo}^{norm}(\Phi)/I_{photo}^{norm}(\Phi=0) =1 $ it can be argued that the three parameters have not changed as a consequence of irradiation. In Fig.~\ref{Fig:photocurrent2} from Ref.~\cite{CentisVignali:2017zpz}, $R$ is presented for a KETEK SiPM operated at -30$^{\circ}$C after neutron irradiation to $\Phi_{eq} = 5\cdot 10^{13}$~cm$^{-2}$. 
Whereas, below breakdown $R$ is still close to unity, it drops rapidly from 1 to 0 above $V_{bd}$. This indicates that one or more parameters are affected at this fluence. 
A possible explanation is that the PDE decreases due to an increase of pixel occupancy. The same argument was also offered by Angelone et al. in Ref.~\cite{Angelone:2010mg} to explain a signal reduction of SiPM irradiated with neutrons to $7\cdot 10^{10}$~cm$^{-2}$.

\begin{figure}[h]
\centering
\includegraphics[width=0.90\linewidth]{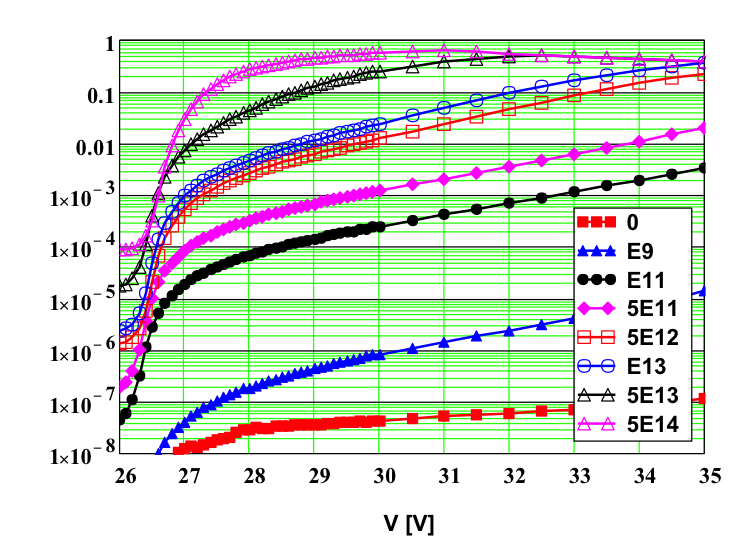}
\caption{The values of $\eta_{DC}$ for SiPM irradiated with neutrons to $\Phi_{eq} = 1\cdot 10^{11} - 5\cdot 10^{14}$~cm$^{-2}$. The measurements are performed at -30~$^{\circ}$C. For fluences larger than $\Phi_{eq} = 10^{12}$~cm$^{-2}$ the pixel occupancy is larger than 1\% at the operating voltage, and therefore a reduction of the effective PDE is expected. In Ref.~\cite{Garutti:2017ipx}, $V_{bd}$ is determined using the inverse logarithmic derivative of the current, and $V_{ex} = V_{bias} -V_{bd}$. }
\label{Fig:eta}
\end{figure}

The pixel occupancy due to dark counts is the probability of a Geiger discharge inside a pixel in a time interval $\Delta t$. Using $\Delta t = \tau = R_q (C_{pix}+C_q) \approx \tau = R_q C_{pix}$ is justified for SiPMs without a large fast component (a large $C_q$ parallel to $R_q$). Then for a device with $N_{pix}$ pixels one obtains~\cite{Garutti:2017ipx}:
\begin{equation}
\eta_{DC} = \frac{I_{dark}}{\Delta V}  \frac{R_q}{N_{pix}} = \frac{\tau ~ECF}{N_{pix}}DCR.
\label{Eq:eta}
\end{equation}
The first equation shows that for $\eta_{DC} =1$ the current over the quenching resistor is equal to the excess bias voltage, so the voltage is always at $V_{bd}$ and the SiPM never recovers. 
For very high DCR, pulses may often occur in a pixel in which the voltage is not fully recovered from a previous discharge, i.e. at an effective lower overvoltage resulting in a smaller gain. This effect may lead to a signal reduction after irradiation. 
In the second part of Eq.~\ref{Eq:eta}, $I_{dark}$ is substituted using Eq.~\ref{eq:idark} and the proportionality to the DCR is made explicit.   
The values of $\eta_{DC}$ measured on SiPMs irradiated with neutrons to $\Phi_{eq} = 1\cdot 10^{11} - 5\cdot 10^{14}$~cm$^{-2}$ are shown in Fig.~\ref{Fig:eta}. The value of $\eta_{DC}$ increases linearly with fluence, and exceeds 1\%  at $\Phi_{eq} > 10^{12}$~cm$^{-2}$, for $\Delta V >$~2~V and $T = -30^{\circ}$C. For a high pixel occupancy a large number of pixels does not recover within the time of the signal and is not available to detect photons, consequently the effective PDE is reduced. It is expected that the signal of the SiPM in response to low intensity of light will decrease at least by the same amount, if no additional effects contribute. 
One should note that the value obtained with this method is an average pixel occupancy and that the deviations between pixels can be very large as already discussed when quoting the work by Barnyakov on digital photon counters,~\cite{Barnyakov:2016bwt}.  

Alternatively, the parameters of irradiated SiPM can be measured from the response to a pulsed light signal (LED or laser) of sufficiently high intensity to be separable from the increased noise. For a pulse of fixed light intensity the integrated SiPM charge follows a distribution with mean $Q$ and variance $\sigma_{Q}^2$ (see Appendix~\ref{appendix}). The $PDE$ is the ratio between mean number of photons photons initiating primary Geiger discharges, $\langle N_{pG} \rangle$, and the average number of photons impinging the photo-detector surface $\langle N_{\gamma} \rangle$. The value of $\langle N_{pG} \rangle$ can be determined in various ways (see Eq.s~\ref{eq:mu1},~\ref{eq:mu2},~\ref{eq:mu3}) each depending on a different assumption. Often used is Eq.~\ref{eq:mu3}, 

\begin{equation}
\label{Eq:PDE}
PDE = \frac{\langle N_{pG} \rangle}{\langle N_{\gamma} \rangle}= \frac{ENF}{\langle N_{\gamma} \rangle}\frac{Q^2}{\sigma_{Q}^2} = ENF~PDE^*,
\end{equation}
where the variance measured in dark conditions is already subtracted from the variance of the light signal: $\sigma_{Q}^2 = \sigma_{Q_{light}}^2 -\sigma_{Q_0}^2$.

In Ref.~\cite{MUSIENKO200987}, results for PDE$^* $ for SiPMs from various producers are compared before and after irradiation with protons to $\Phi_{eq} = 2 \cdot 10^{10}$~cm$^{-2}$. Relative changes in PDE$^*$ are below the 10\% level despite the significant increase of $I_{dark}$ and DCR. 

The effect of high-fluence proton irradiation is reported in Ref.~\cite{Heering:2016lmu}.
In Eq.~\ref{Eq:PDE} the mean number of photo-electrons is directly derived from the amplitude and width of the signal response to a known $N_{\gamma}$. Note that this formula only holds under the assumption of Poisson distributed light intensity, which for instance for laser light of low intensity could become an issue, and for LED pulsed light requires the stability of the pulse generator to be of order 0.1\%. In addition, the light intensity should be low to avoid effects of SiPM non-linearity, and the readout system (preamplifier, ADC, ...) must be linear and have no saturation effects.
In Ref.~\cite{Heering:2016lmu} it is argued that the observed decrease of this quantity indicates a PDE$^* $ decrease by 25\% after $\Phi_{eq} = 2.2 \cdot 10^{14}$~cm$^{-2}$. For the same fluence the signal response to LED light, $Q$ decreases by 50\%. The PDE$^* $ reduction can be explained with the pixel occupancy mechanism, but the additional signal loss must have another origin. Like the photo-current, also the signal is proportional to the product $PDE(\Phi) ~G(\Phi) ~ECF(\Phi)$. To explain the large observed signal reduction either the gain or the $ECF$ must reduce as well. 

A possible additional effect not affecting the pixel occupancy is the decrease of gain. As argued in Ref.~\cite{Heering:2016lmu}, self-heating effects at the pixel level could cause a local breakdown voltage change for the hot pixels. The exact temperature inside the discharging pixel is not easily measurable, but possibly local temperature increase could change the pixel breakdown voltage with respect to the $V_{bd}$ value obtained from I-V measurements. A possible way to further investigate this effect is to compare the three definitions of $\langle N_{pG} \rangle$ given in Appendix~\ref{appendix}, which have different dependencies on the gain (none, first order and quadratic) and therefore may be more sensitive to gain changes. New methods to quantify the gain for irradiated SiPMs would also be beneficial.

\subsection{Annealing effects}
\label{sec:annealing}
Several groups have measured the effect of thermal annealing on irradiated SiPMs. 
Nakamura et al.~\cite{Nakamura09} investigate SiPMs irradiated with neutrons up to $\Phi_{eq} \approx 10^{12}$~cm$^{-2}$, and demonstrate that the single photon detection capability is lost immediately after irradiation, but up to $\Phi_{eq} \approx 10^{10}$~cm$^{-2}$ it recovers after about two months of room temperature annealing. This self-annealing effect is also reported in Ref.~\cite{Andreotti:2013nra, BOHN2009722, SANCHEZMAJOS2009506}.  

Quiang et al.~\cite{QIANG2013234} further investigated the temperature dependent annealing on SiPMs irradiated to $\Phi_{eq} = 3.7\cdot 10^{9}$~cm$^{-2}$. They confirm a self-annealing time constant of about 10 days at room temperature, and show an exponential decrease of the annealing time constant with increasing temperature in the range of -5 / +60~$^{\circ}$C.

\begin{figure}[h]
\centering
\includegraphics[width=0.90\linewidth]{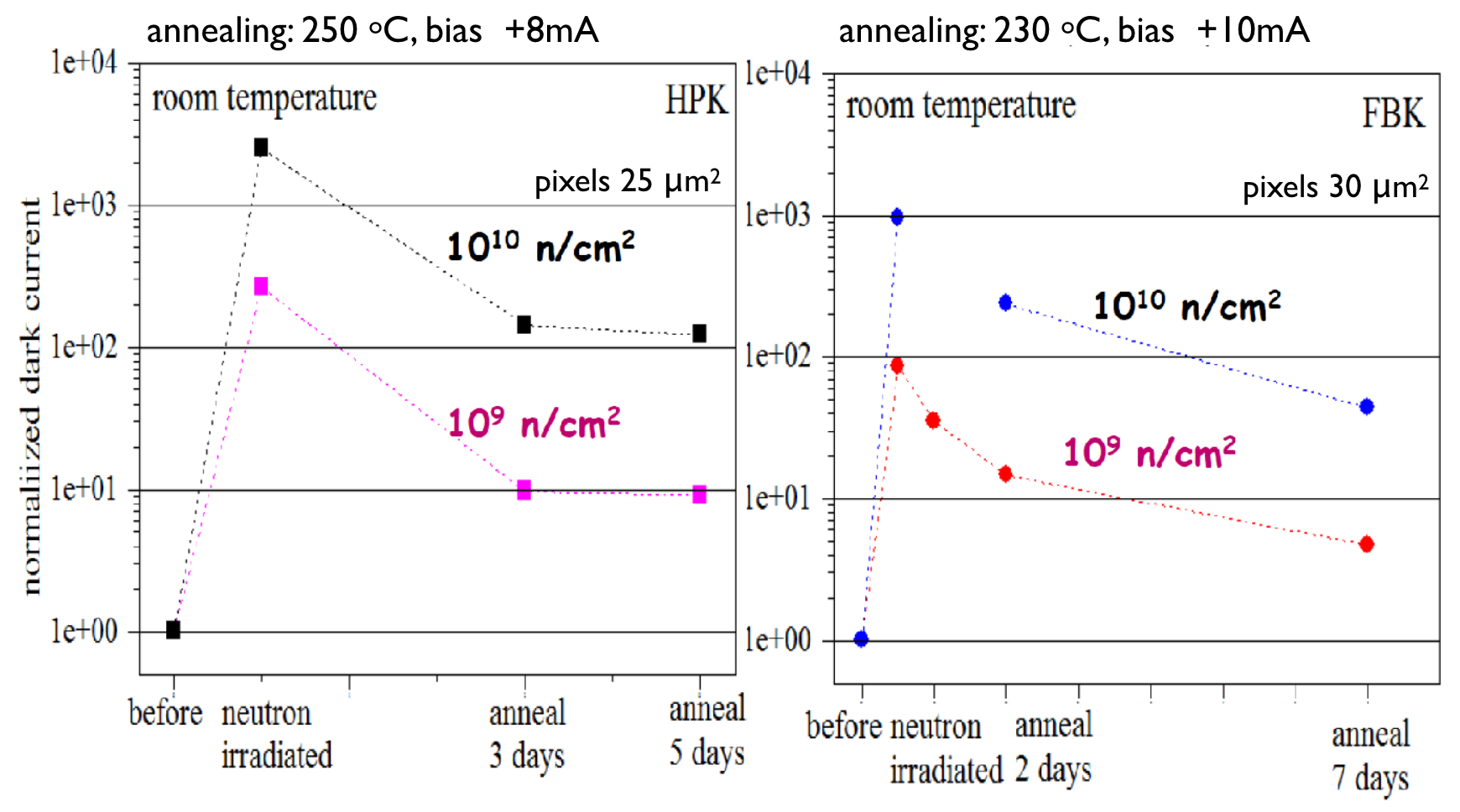}
\caption{Effect of high temperature and forward bias annealing on two types of SiPMs irradiated to $\Phi_{eq} = 10^{9}$~cm$^{-2}$ and $\Phi_{eq} = 10^{10}$~cm$^{-2}$. Left SiPMs from Hamamatsu (HPK) and right from FBK. From Ref.~\cite{Tsang2016}.}
\label{Fig:tsang}
\end{figure}

Tsang has presented results of annealing at +250~$^{\circ}$C, using forward bias with the SiPM current reaching 10~mA,~\cite{Tsang2016}. A remarkable effect of this high temperature annealing is demonstrated in Fig.~\ref{Fig:tsang}. Single photo-electron resolution was recovered after this procedure for devices irradiated up to $\Phi_{eq} = 10^{12}$~cm$^{-2}$ with cooling them to about -50~$^{\circ}$C. Still the dark current before irradiation cannot be completely recovered even after days at the annealing conditions. 

Gotti at al. have recently published a study~\cite{Gotti2018:arxiv} of SiPMs irradiated with neutrons up to $\Phi_{eq} = 10^{14}$~cm$^{-2}$ and demonstrated that single photo-detection with a DCR below 1~kHz is possible after this fluence if  the devices are annealed for several days at 175~$^{\circ}$C, and subsequently operated at cryogenic temperatures (-77~$^{\circ}$C). 


\section{Radiation damage of avalanche photodiodes}
\label{sec:apd}
Avalanche photodiodes (APD) have been established as photo-detectors few years before SiPMs were developed and have been investigated in great detail. Most of the studies of radiation damage in APDs were performed for their application in the electromagnetic calorimeter of the Compact Muon Solenoid (CMS) experiment at the Large Hadron Collider (LHC)~\cite{Chatrchyan:2008aa}.
Analogously to a SiPM, the principle of operation of an APD is based on the conversion of the energy of the absorbed photons into electron-holes pairs in silicon diodes and their multiplication through an avalanche process~\cite{Sze2006}. The diode has a region with electric field of 1.5-2.5$\cdot 10^5$~V/cm (so lower than the field required for a SiPM to reach Geiger mode) where electrons (or holes) are accelerated and acquire a kinetic energy large enough to create additional electron-hole pairs via impact ionization. The breakdown voltage is (in contrast to SiPMs) the maximum voltage that can be applied to an APD. 
The effect of radiation in silicon APDs is very similar to that in silicon diodes operated without gain ~\cite{ANTUNOVIC2005379}. However, there are several important differences related to their operation as photo-detectors with internal gain. It is relevant to give a short review on APDs radiation hard studies, which may prove valuable to interpret results on SiPMs. In the following we report separately on studies performed irradiating APDs with  photons and hadrons.

\subsection{Radiation damage by gammas}
\label{sec:apd-gamma}
The effect of the gamma radiation has been studied in \cite{Kirn:1997}. There it has been established that  APDs, which had a front window made of SiO$_2$, show a decrease of the quantum efficiency in the short wavelength region after gamma irradiation, while the photosensors, with a Si$_3$N$_4$ window, show no change in the quantum efficiency after exposure to a dose of 55~kGy. 
More than 3000 CMS APDs were irradiated with $^{60}$Co source up to 2.5~kGy gamma dose~\cite{ANTUNOVIC2005379}. A dark current increase of $\sim$150-500~nA was measured after irradiation at the operating voltage (corresponding to $G = 50$). The origin of this current was associated to damage in the APD dielectric layer, as it had only a weak dependence on the APD gain. 
The maximum operating voltage, corresponding to the breakdown voltage was in the range of 430-440 V before irradiation, and was reduced by 4-30 V for approximately 0.7\% of the irradiated APDs. The other APDs presented no effect of breakdown voltage decrease.

\subsection{Radiation damage by hadrons}
\subsubsection{Dark current, QE and gain}
\label{sec:apd-DC}
An extensive series of studies of the APD resistance to hadron radiation has been carried out by different groups participating in the CMS collaboration. These studies have been performed utilizing reactor neutrons, a spallation neutron source, a $^{252}$Cf source, and a proton beam (see Ref.~\cite{ANTUNOVIC2005379} and Refs. therein).

\begin{figure}[h]
\centering
\includegraphics[width=0.7\linewidth]{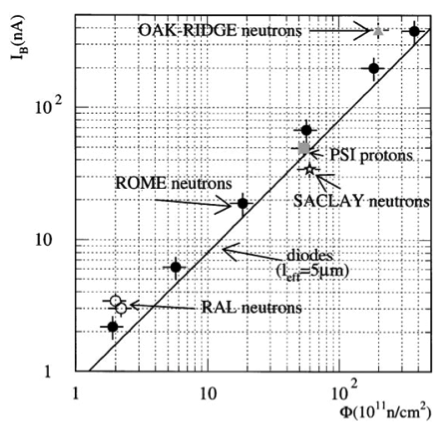}
\caption{Radiation-induced bulk current versus fluence for one of the Hamamatsu APDs (25 mm$^2$ in area). The measurements are taken at 18$^{\circ}$C and  2 days after the irradiation~\cite{BACCARO1999206}. }
\label{Fig:APD2}
\end{figure}

Fig.~\ref{Fig:APD2} shows the increase in the bulk current as a function of the neutron fluence~\cite{BACCARO1999206}. The parameter "bulk current" or "generation bulk current" was introduced in this paper because it was found that the ratio of dark current and gain is almost constant in a wide range of gains ($G >$ 10). The bulk current is defined as the dark current divided by the APD gain ($G =$ 50 in this case). It can be noticed that the bulk current grows linearly with the fluence, as for diodes.

When trying to compare the damage parameter $\alpha$ for APDs and diodes (see solid line in Fig.~\ref{Fig:APD2}), it was observed that the volume of the APD contributing to the dark current generation is smaller than the geometrical volume obtained multiplying the active area by the depletion depth (for Hamamatsu APDs: A = 25~mm$^2$, $L_{dep}$ = 50~$\mu m$). This is because the carriers generated throughout the depletion region in a device with multiplication do not all contribute the same to the total dark current. The correct formalism to describe the noise generation in a device with multiplication gain was developed by McIntyre in Ref.~\cite{McIntyre:1966}. In Ref.~\cite{Musienko:2000jj} the concept of an effective thickness of the APD, $L_{eff}$ was introduced for simplicity. It is found that the contribution to the bulk current is generated by a region with $L_{eff} \approx$ 5~$\mu m$, compared to the depletion thickness of $L_{dep}$ = 50~$\mu m$. The same effect is also relevant when discussing the region contributing to the dark current of a SiPMs.

\begin{figure}[h]
\centering
\includegraphics[width=0.7\linewidth]{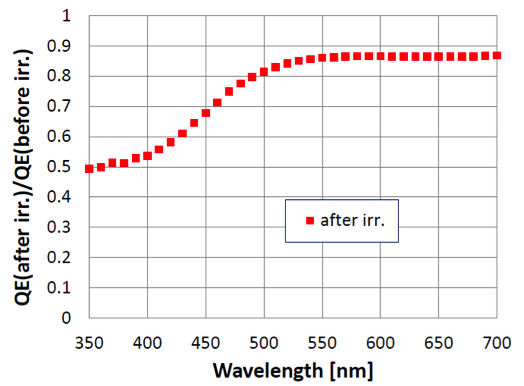}
\caption{Quantum efficiency losses vs. wavelength after irradiation (G~=~19, T~=~25~$^{\circ}$C), after irradiation to $\Phi_{eq}=5\cdot 10^{14}$~cm$^{-2}$ and several years of annealing at room temperature. Plot from Ref.~\cite{Musienko:2015fiz}.}
\label{Fig:APD5}
\end{figure}

\begin{figure}[h]
\centering
\includegraphics[width=0.7\linewidth]{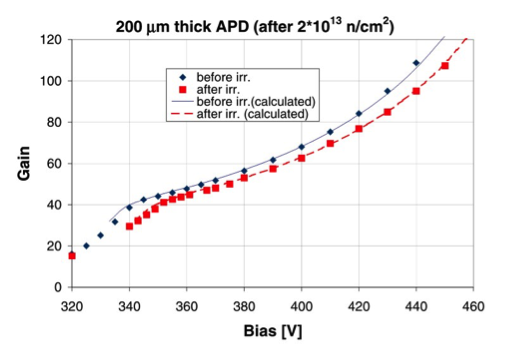}
\caption{Gain as a function of bias voltage for the 196 um EG\&G APD, measured before and after irradiation.}
\label{Fig:APD3}
\end{figure}

In Ref.~\cite{Musienko:2000jj} two APDs from EG\&G with different depletion region thicknesses, 196 and 243 $\mu$m, and one APD from Hamamatsu, with depletion region thickness, $\sim$50~$\mu$m, were irradiated to $\Phi_{eq}=$ 2.33$\cdot 10^{13}$~cm$^{-2}$. The quantum efficiencies of the investigated EG\&G APDs remained unchanged over a wide range of wavelengths. The Hamamatsu APD showed a significant decrease of quantum efficiency (by 10-40 \%), especially for short wavelengths, as shown in Fig.~\ref{Fig:APD5}.
In Ref.~\cite{Musienko:2000jj} it was observed that the curves of gain as a function of bias voltage for the EG\&G APDs after irradiation are shifted compared to those before irradiation (see Fig.~\ref{Fig:APD3}). This effect is related to a change of the doping concentration inside the depletion region of the APD due to the creation of acceptor-like states~\cite{Musienko:2000im}, that caused the APD breakdown voltage increase. The gain as a function of bias voltage for the Hamamatsu APD remained unchanged after irradiation, probably due to a thin depletion region ($\sim$50~$\mu$m).

The CMS ECAL APDs (produced by Hamamatsu) were irradiated up to $\Phi_{eq}=5\cdot 10^{14}$~cm$^{-2}$~\cite{Musienko:2015fiz}. Dark  current and gain vs. bias dependences were measured at T~=~25~$^{\circ}$C, 15~$^{\circ}$C and 5~$^{\circ}$C after several years of annealing at room temperature. The gain vs. bias and QE vs. wavelength dependences were compared with that of non-irradiated APDs. The main results of the study are:
\begin{itemize}
\item 	APDs are still operational as light detectors with $G>15$ also without cooling;

\item The dark current increases linearly with fluence. After $\Phi_{eq}=5\cdot 10^{14}$~cm$^{-2}$, and for $G=10$ and T~=~25~$^{\circ}$C, it is $I_{dark} \approx$ 80~$\mu$A;

\item The dark current can be significantly decreased by reduction of temperature ($\sim$2.2 times per 10~$^{\circ}$C, consistent with the temperature dependence of leakage current in diodes, in the absence of strong fields) $-$ For SiPMs the reduction of dark current with temperature is weaker.

\item QE losses of 20-50\% were found for the 350-500 nm wavelength range (see Fig.~\ref{Fig:APD5}) $-$ Similar measurements are not available for SiPMs;

\item The gain at a fixed voltage was reduced due to doping profile change, caused by creation of  acceptor-like states in the APD depleted region. The problem can be solved by $\sim$30~V bias voltage increase $-$ This effect is qualitatively consistent with the shift in $V_{bd}$ observed in SiPMs; 

\end{itemize}

\subsubsection{Annealing studies}
\label{sec:apd-annealing}
The recovery of the dark current can be described~\cite{BACCARO1999206} as a sum of exponentials, each exponential being attributed to the recovery of one defect with its proper recovery-time $\tau_i$ and its weight $g_i$:
\begin{equation}
I_{dark}^{irr}(t) = I_{dark}^{irr}(0) \sum{g_i e^{-\frac{t}{\tau_i}}}
\label{Eq:APD3}
\end{equation}

\begin{figure}[h]
\centering
\includegraphics[width=0.65\linewidth]{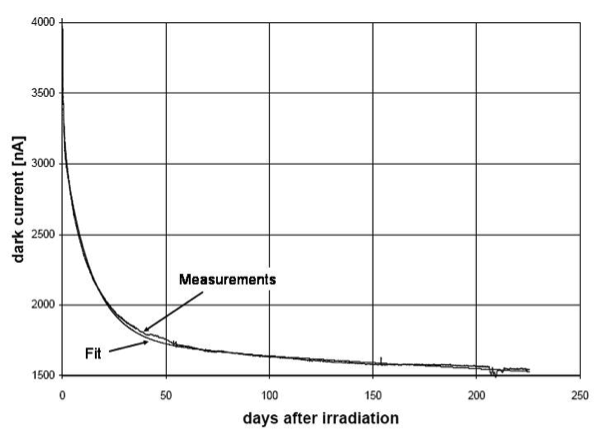}
\caption{Room temperature annealing of the dark current of one of the Hamamatsu APDs irradiated with 2.7$\cdot 10^{12}$ (E=72 MeV) protons/cm$^{2}$ at PSI~\cite{CMSECAL:1997}.}
\label{Fig:APD4}
\end{figure}

Fig.~\ref{Fig:APD4} shows the recovery of the dark current of an APD irradiated to $\Phi=$2.7$\cdot 10^{12}$ (E=72 MeV) protons/cm$^{2}$. The measurements were done at $G=50$ and at room temperature. The measurements started 10 min after irradiation. Table~\ref{Tab:tau} shows the decay-times and the weights of the various components fitted from these data points. 

\begin{table}
\label{Tab:tau}
\centering
\begin{tabular}{|c|c|}
\hline
$g_i$ & $\tau_i$\\
\hline
0.21 & 576 min \\
0.36 & 13 days \\
0.17 & 600 days \\
0.26 & infinite \\
\hline
\end{tabular}
\caption{Dark current annealing decay times and their weights for Hamamatsu APDs irradiated with 2.7$\cdot 10^{12}$ (E=72 MeV) protons/cm$^{2}$ at PSI~\cite{CMSECAL:1997}. The values are extracted by a fit to Fig.~\ref{Fig:APD4} using Eq.~\ref{Eq:APD3}.
}
\end{table}

Several other APDs have been irradiated with neutrons. They show similar annealing behaviors (see Ref.~\cite{BACCARO1999206}). 

\section{Approaches to develop radiation harder SiPMs}
\label{sec:outlook}
There are several factors limiting the operation of SiPMs in high radiation environment:
\begin{itemize}
\item dark noise increase due to deep level defects produced by radiation in silicon bulk;
\item 	high cell occupancy due to high generation-recombination rate and limited cell recovery time causes a reduction of SiPM gain and PDE;
\item 	large leakage current results in increase of the p-n junction temperature and high power consumption;
\item 	increase of the SiPM breakdown voltage due to induced changes in doping concentration;
\item 	reduction of the SiPM photon detection efficiency due to damage in dielectric-silicon interface and charge trapping in non-depleted entrance layer;
\end{itemize}
Next we discuss approaches to overcome these limitations.\\

\noindent\textbf{Dark noise reduction}\\
The main problem for the optimization of radiation tolerance of the SiPMs is presented by the increase of DCR. The main sources of the DCR increase are:
\begin{itemize}
\item surface current reaching the multiplication region;
\item diffusion from the non-depleted bulk;
\item defect-induced generation in the depletion region. 
\end{itemize}
Dark noise generation in SiPMs can be reduced by 
\begin{itemize}
\item avoiding that surface current reaches the multiplication region;
\item reducing diffusion from the non-depleted bulk;
\item optimizing the field in the depletion region, "field shaping".
\end{itemize}
This last point can be achieved by:
\begin{itemize}
\item reducing thickness of the depleted region; 
\item reducing trap-assisted tunnelling (reducing peak electric field in the SiPM p-n-junction). 
\end{itemize}
Unfortunately these two approaches contradict each other: a reduction of electric field requires wider avalanche and depletion regions. A special R\&D has to be performed to find an optimum (compromise) between these two approaches. \\

\noindent\textbf{Cell occupancy reduction}\\
The cell occupancy reduction can be achieved by reducing the cell active volume (smaller cell size) and cell recovery time. However SiPMs with high cell density have large non-sensitive zones (cell separation regions, area occupied by polysilicon quenching resistors which are almost non-transparent). The solution of this problem can be very thin trenches ($<$0.5~$\mu$m) separating cells and use of metal film resistors with high transparency to visible light. Smaller cell size results also in smaller cell capacitance and reduction of the recovery time. Reduction of quenching resistor value is limited. Few hundreds k$\Omega$ resistors are required to quench the cell discharge and allow stable SiPM operation in a reasonable range of over-voltages (2-3~V).
The reduction of the SiPM cell size and capacitance has another positive outcome, namely the reduction of the SiPM dark noise, or leakage current (mainly because of gain reduction). This is also linked to a reduction of power consumption. 
Related to the need of short recovery time is another important aspect, namely that of including the signal shaping (fast recovery time, one vs two time component, ...) in the optimization process. So far this aspect has not been covered in literature. \\

\noindent\textbf{Breakdown voltage increase minimization}\\
The SiPM $V_{bd}$ increase with irradiation strongly depends on the thickness of the depletion region. It can be reduced by reducing this thickness. SiPMs with low $V_{bd}$ value have significantly smaller $V_{bd}$ increase after hadron irradiation \cite{Andreotti:2013nra,Heering:2016lmu}.\\

\noindent\textbf{Reduction of the damage in SiPM entrance window}\\
The SiPM entrance window material has to be chosen appropriately for radiation hard applications, i.e. avoiding materials with hydrogen or boron content. 
The thickness of the SiO$_2$ layer has to be properly adjusted taking into account surface effects. The thickness of non-depleted region near the SiPM entrance window has to be minimized to reduce nuclear effects in its volume.\\

Detailed and possibly realistic simulations of SiPMs should be performed in order to optimize the SiPM designs for specific applications.
For this purpose information is still missing:
\begin{itemize}
\item Fluence dependence of relevant parameters: mobility and multiplication coefficient as function of doping; field enhanced generation of traps,
\item more systematic measurements of SiPM parameters,
\item microscopic measurements on defects (can defect engineering help?). 
\end{itemize}
The outcome of the optimization will depend on the targeted wavelength and the expected operation temperature for the SiPM.

\section{Conclusions}
This review is an attempt to summarize the current knowledge of radiation damage of SiPMs, a complex multi-parameter problem that recently has attracted large attention in the SiPMs users community. The main issue with irradiated SiPMs is the increase of dark count rate, which at some point also affects other SiPM parameters. Due to the increased DCR, the single photo-electron separation from noise is lost already at relatively low fluences $\Phi_{eq} \sim 10^{10}$~cm$^{-2}$. This limit depends on many factors related to the SiPM design and the operation conditions, so it should be tested for each specific application. The DCR also affects the pixel occupancy, which if no longer negligible yields to $PDE$ reduction. 

Once single photo-electrons can no longer be resolved, many characterization methods used for the investigation of non-irradiated SiPM parameters fail. Some recently developed characterization methods are presented in this review. We tried to spell our the assumptions made by each method. They all still need further testing and validation on larger sets of data. It is reported that SiPM parameters like $R_q$, $C_{pix}$, $V_{bd}$, $G$, $ECF$, $ENF$, $PDE$ for $\Phi_{eq} < 10^{12}$~cm$^{-2}$, whereas above this fluence changes are observed. Again this limit is SiPM design and temperature dependent. A proper quantification of the fluence dependence of the individual parameters requires further dedicated experiments. 
Also the annealing state of the SiPM is important to define limits on the operation range. Annealing studies indicate the possibility to recover single photo-electron resolution at least for operation at low/cryogenic temperatures. 

An optimization of the SiPM design for operation in high radiation environment is not yet available and lacks still dedicated experimental input. 
Further effort in this field is strongy encouraged.

\appendix
\section{Statistical treatment of SiPM signal}
\label{appendix}
The statistical treatment of SiPM signals is presented in several papers, notable are among other Ref.~\cite{Vinogradov:2011vr,Arosio:2015sra,Chmill:2016ghf}. 

Given a Poisson distribution of impinging photons on the SiPM, its response does not follow a Poisson distribution due to its correlated noise. A generalized Poisson formalism is used often to account for in-time correlated signal generation (prompt cross-talk). But SiPMs are also affected by late correlated noise such as delayed cross-talk and after-pulses. We report here some useful definitions related to the treatment of SiPM signals and relevant for the formalism in this paper. 

The charge distribution of a SiPM integrated in a time $t_{gate}$, is characterized by the first moment $\langle Q \rangle$ and the second central moment or variance, $\sigma_Q^2 = Var_Q = E[(Q-\langle Q \rangle)^2]$. In the following the variables $\langle Q \rangle$, and $\sigma_Q^2$ will be used for the mean and variance of the charge distribution. Another often used distribution to characterize SiPMs is that of the number of measured Geiger discharges with mean $\langle N_{pe} \rangle$ and the variance $\sigma_{pe}^2$. 
The distributions of charge and number of measured Geiger discharges, or photo-electrons, are linked by the relations:
\begin{equation}
\langle Q \rangle = q_0 ~G ~\langle N_{pe} \rangle,
\label{eq:multimean0}
\end{equation}
\begin{equation}
\sigma_Q^2 = q_0^2 ~G^2 ~\sigma_{pe}^2.
\label{eq:multivar0}
\end{equation}

The definition of the Excess Charge Factor (ECF) (Eq.~\ref{eq:ecf}) relates the mean charge to that of the primary Geiger discharges (Poisson distributed): 
\begin{equation}
ECF =\frac{\langle Q \rangle}{\langle Q_P \rangle} =\frac{\langle N_{pe} \rangle}{\langle N_{pG} \rangle}, 
\label{eq:eqf}
\end{equation}
with $\langle N_{pG} \rangle = \mu$ being the mean number of primary Geiger discharges with $\sigma_{pG} = \sqrt{\langle N_{pG} \rangle}$ .
The Excess Noise Factor (ENF) relates the variances of the two distributions (Eq.~\ref{eq:enf}):
\begin{equation}
ENF=\frac{\sigma_Q^2}{\langle Q \rangle ^2} \frac{\langle Q^2_P \rangle}{(\sigma_Q^2)_P}
   =\frac{1}{ECF^2} \frac{\sigma_Q^2}{(\sigma_Q^2)_P} 
   =\frac{1}{ECF^2} \frac{\sigma_{pe}^2}{\langle N_{pG} \rangle}.
   \label{eq:enf1}
\end{equation}
We note that for the Generalized Poisson distribution $GP(k;\mu, \lambda) = (\mu \cdot (\mu + k \cdot \lambda)^{k-1} \cdot e^{-(\mu + k\cdot\lambda)}/k!$, $ECF = ENF = 1/(1 -\lambda)$~\cite{Vinogradov:2011vr}. 

Substituting Eq.~\ref{eq:eqf} in Eq.~\ref{eq:multimean0}, and Eq.~\ref{eq:enf1} in Eq.~\ref{eq:multivar0} one gets:






\begin{equation}
\langle Q \rangle=   q_0 ~G ~ECF ~\langle N_{pG} \rangle,
\label{eq:multimean}
\end{equation}
\begin{equation}
\sigma_Q^2  = q_0^2 ~G^2 ~ENF ~ECF^2 ~\langle N_{pG} \rangle.
\label{eq:multivar}
\end{equation}


From Eq.~\ref{eq:multimean} and Eq.~\ref{eq:multivar} the gain factor (G) is obtained as:
\begin{equation}
G = \frac{1}{q_0~ ENF~ECF}\frac{\sigma_Q^2}{\langle Q \rangle}.
\label{eq:gain2}
\end{equation}
Eq.~\ref{eq:gain2} can be used when a SiPM does not provide single photo-electron resolution, as for instance after radiation damage. \\ 

For the analysis of light signals, the mean number of primary Geiger avalanches, $\langle N_{pG} \rangle$ can be determined in two different ways using either Eq.~\ref{eq:multimean} or Eq.~\ref{eq:multivar}:
\begin{equation}
\langle N_{pG} \rangle_1 = \frac{\langle Q \rangle}{q_0 ~G ~ECF} 
\label{eq:mu1}
\end{equation}
\begin{equation}
\langle N_{pG} \rangle_2= \frac{\sigma_Q^2} {q_0^2 ~G^2 ~ENF ~ECF^2 },
\label{eq:mu2}
\end{equation}
Combining Eq.~\ref{eq:eqf} in Eq.~\ref{eq:enf1} a third independent definition is:
\begin{equation}
\langle N_{pG} \rangle_3= ENF \frac{\langle Q \rangle ^2}{\sigma_Q^2} = ENF \frac{\langle N_{pe} \rangle ^2}{\sigma_{pe}^2},
\label{eq:mu3}
\end{equation}
where mean and variance can be taken consistently from the distribution of charge, number of photo-electrons or even ADC channels with no need of further calibration, since additional conversion factors cancel in the ratio. \\

\bibliographystyle{elsarticle-num}
\bibliography{bibliography}

\begin{thebibliography}{10}
\expandafter\ifx\csname url\endcsname\relax
  \def\url#1{\texttt{#1}}\fi
\expandafter\ifx\csname urlprefix\endcsname\relax\def\urlprefix{URL }\fi
\expandafter\ifx\csname href\endcsname\relax
  \def\href#1#2{#2} \def\path#1{#1}\fi

\bibitem{Piemonte:thisVolume}
C.~Piemonte, A.~Cola, Overview on the main parameters and technology of silicon
  photo-multipliers, this Special Issue on SiPMs.

\bibitem{LUTZ1999}
G.~Lutz, Semiconductor radiation detectors - device physics, Springer Verlag
  Berlin/Heidelberg.

\bibitem{LINDSTROM200330}
G.~Lindstrom, Radiation damage in silicon detectors, Nucl. Instrum. Meth.
  A512~(1) (2003) 30 -- 43.
\newblock \href
  {http://dx.doi.org/https://doi.org/10.1016/S0168-9002(03)01874-6}
  {\path{doi:https://doi.org/10.1016/S0168-9002(03)01874-6}}.

\bibitem{Moll:1999kv}
M.~Moll, {Radiation damage in silicon particle detectors: Microscopic defects
  and macroscopic properties}, Ph.D. thesis, Hamburg U. (1999).

\bibitem{Haitz2006}
R.~H. Haitz, Mechanisms contributing to the noise pulse rate of avalanche
  diodes, Journal of Applied Physics 36~(10) (2006) 3123 -- 3131.
\newblock \href {http://dx.doi.org/https://doi.org/10.1063/1.1702936}
  {\path{doi:https://doi.org/10.1063/1.1702936}}.

\bibitem{Pagano2012}
R.~Pagano, D.~Corso, S.~Lombardo, G.~Valvo, D.~N. Sanlippo, G.~Fallica,
  S.~Libertino, Dark current in silicon photomultiplier pixels: Data and model,
  IEEE Transactions on Nuclear Science 59~(2) (2012) 2410 -- 2416.
\newblock \href {http://dx.doi.org/https://doi.org/10.1109/TED.2012.2205689}
  {\path{doi:https://doi.org/10.1109/TED.2012.2205689}}.

\bibitem{Vincent1979}
G.~Vincent, A.~Chantre, D.~Bois, Electric field effect on the thermal emission
  of traps in semiconductor junctions, Appl. Phys. 50~(8) (1979) 5484--5487.
\newblock \href {http://dx.doi.org/https://doi.org/10.1063/1.326601}
  {\path{doi:https://doi.org/10.1063/1.326601}}.

\bibitem{Hurkx1992}
G.~Hurkx, D.~Klaassen, M.~Knuvers, A new recombination model for device
  simulation including tunneling, IEEE Transactions on Nuclear Science 39~(2)
  (1992) 331 -- 338.
\newblock \href {http://dx.doi.org/https://doi.org/10.1109/16.121690}
  {\path{doi:https://doi.org/10.1109/16.121690}}.

\bibitem{Grove1967}
A.~S. Grove, Physics and technology of semiconductor devices, John Wiley \&
  Sons.

\bibitem{Chilin:2013}
A.~Chilingarov, {Temperature dependence of the current generated in Si bulk},
  JINST 8~(10)  P10003.

\bibitem{Hurkx1992:field}
G.~Hurkx, H.~de~Graaff, W.~Kloosterman, M.~Knuvers, A new analytical diode
  model including tunneling and avalanche breakdown, IEEE Transactions on
  Nuclear Science 39~(9) (1992) 2090 -- 2098.
\newblock \href {http://dx.doi.org/https://doi.org/10.1109/16.155882}
  {\path{doi:https://doi.org/10.1109/16.155882}}.

\bibitem{Acerbi:2016ikf}
F.~Acerbi, et~al., {Cryogenic Characterization of FBK HD Near-UV Sensitive
  SiPMs}, IEEE Transactions Electron Devices 64~(2) (2016) 521--526.
\newblock \href {http://arxiv.org/abs/1610.01915} {\path{arXiv:1610.01915}},
  \href {http://dx.doi.org/10.1109/TED.2016.2641586}
  {\path{doi:10.1109/TED.2016.2641586}}.

\bibitem{Barnaby2006}
H.~J. Barnaby, Total-ionizing-dose effects in modern cmos technologies, IEEE
  Transactions on Nuclear Science 53~(6) (2006) 3103 -- 3121.
\newblock \href {http://dx.doi.org/https://doi.org/10.1109/TNS.2006.885952}
  {\path{doi:https://doi.org/10.1109/TNS.2006.885952}}.

\bibitem{Oldham1999}
T.~Oldham, Ionizing radiation effects in mos oxides, World Scientic Publishing
  Co.

\bibitem{Klanner2013}
R.~Klanner, E.~Fretwurst, I.~Pintilie, J.~Schwandt, J.~Zhang, Study of
  high-dose x-ray radiation damage of silicon sensors, Nucl. Instrum. Meth.
  A732 (2013) 117 -- 121.
\newblock \href {http://dx.doi.org/doi:10.1016/j.nima.2013.05.131}
  {\path{doi:doi:10.1016/j.nima.2013.05.131}}.

\bibitem{Zhang:2013}
J.~Zhang, {X-ray radiation damage studies and design of a silicon pixel sensor
  for science at the XFEL}, Ph.D. thesis, Hamburg U. (2013).

\bibitem{Lindstrom:1999mw}
G.~Lindstrom, M.~Moll, E.~Fretwurst, {Radiation hardness of silicon detectors:
  A challenge from high-energy physics}, Nucl. Instrum. Meth. A426 (1999)
  1--15.
\newblock \href {http://dx.doi.org/10.1016/S0168-9002(98)01462-4}
  {\path{doi:10.1016/S0168-9002(98)01462-4}}.

\bibitem{Moll:1999nh}
M.~Moll, E.~Fretwurst, G.~Lindstrom, {Leakage current of hadron irradiated
  silicon detectors - material dependence}, Nucl. Instrum. Meth. A426 (1999)
  87--93.
\newblock \href {http://dx.doi.org/10.1016/S0168-9002(98)01475-2}
  {\path{doi:10.1016/S0168-9002(98)01475-2}}.

\bibitem{RADECSDonegani}
E.~M. Donegani, E.~Fretwurst, E.~Garutti, {Defect spectroscopy of
  proton-irradiated thin p-type silicon sensors}, {2016 16th European
  Conference on Radiation and Its Effects on Components and Systems (RADECS).
  }\href {http://dx.doi.org/10.1109/RADECS.2016.8093113}
  {\path{doi:10.1109/RADECS.2016.8093113}}.

\bibitem{DONEGANI201815}
Study of point- and cluster-defects in radiation-damaged silicon, Nucl.
  Instrum. Meth. A898  15 -- 23.
\newblock \href {http://dx.doi.org/https://doi.org/10.1016/j.nima.2018.04.051}
  {\path{doi:https://doi.org/10.1016/j.nima.2018.04.051}}.

\bibitem{Barnyakov:2016bwt}
M.~{\relax Yu}. Barnyakov, T.~Frach, S.~A. Kononov, I.~Kuyanov, V.~Prisekin,
  {Radiation hardness test of the Philips Digital Photon Counter with proton
  beam}, Nucl. Instrum. Meth. A824 (2016) 83--84.
\newblock \href {http://dx.doi.org/10.1016/j.nima.2015.10.098}
  {\path{doi:10.1016/j.nima.2015.10.098}}.

\bibitem{Newman}
R.~Newman, {Visible Light from a Silicon p-n Junction}, Phys. Rev. 100 (1955)
  700--703.

\bibitem{Lacaita}
A.~L. Lacaita, et~al., {On the bremsstrahlung origin of hot-carrier-induced
  photons in silicon devices}, IEEE Transactions Electron Devices 40.3 (1993)
  577--582.

\bibitem{Engelmann:phd}
E.~Engelmann, {Dark count rate of silicon photo-multipliers}, Ph.D. thesis,
  Universitaet der Bundeswehr Muenchen (2018).

\bibitem{Garutti:2017ipx}
E.~Garutti, R.~Klanner, D.~Lomidze, J.~Schwandt, M.~Zvolsky, {Characterisation
  of highly radiation-damaged SiPMs using current measurements},\href
  {http://arxiv.org/abs/1709.05226} {\path{arXiv:1709.05226}}.

\bibitem{Tsang2016}
T.~Tsang, T.~Rao, S.~Stoll, C.~Woody, {Neutron radiation damage and recovery
  studies of SiPMs}, JINST 11~(12) (2016) P12002.

\bibitem{Klanner:thisVolume}
R.~Klanner, {Characterization of SiPMs}, this Special Issue on SiPMs.

\bibitem{Garutti2016:IEEE}
M.~Centis~Vignali, V.~Chmill, E.~Garutti, R.~Klanner, M.~Nitschke, J.~Schwandt,
  S.~Sonder, {Neutron induced radiation damage of KETEK SiPMs}, IEEE explore
  NSS/MIC/RDNT.
\newblock \href {http://dx.doi.org/10.1109/NSSMIC.2016.8069733}
  {\path{doi:10.1109/NSSMIC.2016.8069733}}.

\bibitem{Matsubara:2006zz}
T.~Matsubara, H.~Tanaka, K.~Nitta, M.~Kuze, {Radiation damage of MPPC by
  gamma-ray irradiation with Co-60}, PoS PD07 (2007) 032.

\bibitem{RenkerP04004}
D.~Renker, E.~Lorenz, Advances in solid state photon detectors, JINST 4~(04)
  (2009) P04004.

\bibitem{Pagano:2014bua}
R.~Pagano, S.~Lombardo, F.~Palumbo, D.~Sanfilippo, G.~Valvo, G.~Fallica,
  S.~Libertino, {Radiation hardness of silicon photomultipliers under 60 Co
  γ-ray irradiation}, Nucl. Instrum. Meth. A767 (2014) 347--352.
\newblock \href {http://dx.doi.org/10.1016/j.nima.2014.08.028}
  {\path{doi:10.1016/j.nima.2014.08.028}}.

\bibitem{Sudo2009}
Y.~Sudo, Study of the multi pixel photon counter for the ilc scintillator-strip
  calorimeter, PoS PD09 (2009) 005.

\bibitem{QIANG2013234}
Y.~Qiang, C.~Zorn, F.~Barbosa, E.~Smith, Radiation hardness tests of sipms for
  the jlab hall d barrel calorimeter, Nucl. Instrum. Meth. A698 (2013) 234 --
  241.
\newblock \href {http://dx.doi.org/https://doi.org/10.1016/j.nima.2012.10.015}
  {\path{doi:https://doi.org/10.1016/j.nima.2012.10.015}}.

\bibitem{SANCHEZMAJOS2009506}
S.~Sanchez~Majos, P.~Achenbach, C.~Ayerbe~Gayoso, J.~Bernauer, R.~Bahm,
  M.~Distler, M.~Gomez Rodriguez de~la Paz, H.~Merkel, U.~Mueller,
  L.~Nungesser, J.~Pochodzalla, B.~Schlimme, T.~Walcher, M.~Weinriefer,
  C.~Yoon, Noise and radiation damage in silicon photomultipliers exposed to
  electromagnetic and hadronic radiation, Nucl. Instrum. Meth. A602~(2) (2009)
  506 -- 510.
\newblock \href {http://dx.doi.org/https://doi.org/10.1016/j.nima.2009.01.176}
  {\path{doi:https://doi.org/10.1016/j.nima.2009.01.176}}.

\bibitem{MUSIENKO2007433}
{\relax Yu}.~Musienko, D.~Renker, S.~Reucroft, R.~Scheuermann, A.~Stoykov,
  J.~Swain, Radiation damage studies of multipixel geiger-mode avalanche
  photodiodes, Nucl. Instrum. Meth. A581~(1) (2007) 433 -- 437.
\newblock \href {http://dx.doi.org/https://doi.org/10.1016/j.nima.2007.08.021}
  {\path{doi:https://doi.org/10.1016/j.nima.2007.08.021}}.

\bibitem{Nakamura09}
I.~Nakamura, Radiation damage of pixelated photon detector by neutron
  irradiation, Nucl. Instrum. Meth. A610 (2009) 110--113.
\newblock \href {http://dx.doi.org/10.1016/j.nima.2009.05.086}
  {\path{doi:10.1016/j.nima.2009.05.086}}.

\bibitem{Andreotti:2013nra}
M.~Andreotti, et~al., {Silicon photo-multiplier radiation hardness tests with a
  white neutron beam}, in: {Proceedings, 3rd International Conference on
  Advancements in Nuclear Instrumentation Measurement Methods and their
  Applications (ANIMMA 2013): Marseille, France, June 23-27, 2013}, 2013.
\newblock \href {http://dx.doi.org/10.1109/ANIMMA.2013.6728033}
  {\path{doi:10.1109/ANIMMA.2013.6728033}}.

\bibitem{Musienko:2015lia}
{\relax Yu}.~Musienko, A.~Heering, R.~Ruchti, M.~Wayne, A.~Karneyeu,
  V.~Postoev, {Radiation damage studies of silicon photomultipliers for the CMS
  HCAL phase I upgrade}, Nucl. Instrum. Meth. A787 (2015) 319--322.
\newblock \href {http://dx.doi.org/10.1016/j.nima.2015.01.012}
  {\path{doi:10.1016/j.nima.2015.01.012}}.

\bibitem{Heering:2016lmu}
A.~Heering, {\relax Yu}.~Musienko, R.~Ruchti, M.~Wayne, A.~Karneyeu,
  V.~Postoev, {Effects of very high radiation on SiPMs}, Nucl. Instrum. Meth.
  A824 (2016) 111--114.
\newblock \href {http://dx.doi.org/10.1016/j.nima.2015.11.037}
  {\path{doi:10.1016/j.nima.2015.11.037}}.

\bibitem{CentisVignali:2017zpz}
M.~Centis~Vignali, E.~Garutti, R.~Klanner, D.~Lomidze, J.~Schwandt, {Neutron
  irradiation effect on SiPMs up to $\Phi_{neq}$ = 5 $\times$ 10$^{14}$
  cm$^{-2}$},\href {http://arxiv.org/abs/1709.04648} {\path{arXiv:1709.04648}}.

\bibitem{Musienko:2017znn}
{\relax Yu}.~Musienko, A.~Heering, R.~Ruchti, M.~Wayne, {\relax Yu}.~Andreev,
  A.~Karneyeu, V.~Postoev, {Radiation damage in silicon photomultipliers
  exposed to neutron radiation}, JINST 12~(07) (2017) C07030.
\newblock \href {http://dx.doi.org/10.1088/1748-0221/12/07/C07030}
  {\path{doi:10.1088/1748-0221/12/07/C07030}}.

\bibitem{Cattaneo17}
P.~Cattaneo, T.~Cervi, A.~Menegolli, M.~Oddone, M.~Prata, M.~Prata,
  M.~Rossella, Radiation hardness tests with neutron flux on different silicon
  photomultiplier devices, JINST 12~(07) (2017) C07012.

\bibitem{Matsumura:2009he}
T.~Matsumura, et~al., {Effects of radiation damage caused by proton irradiation
  on Multi-Pixel Photon Counters (MPPCs)}, Nucl. Instrum. Meth. A603 (2009)
  301--308.
\newblock \href {http://arxiv.org/abs/0901.2430} {\path{arXiv:0901.2430}},
  \href {http://dx.doi.org/10.1016/j.nima.2009.02.022}
  {\path{doi:10.1016/j.nima.2009.02.022}}.

\bibitem{DANILOV2009183}
M.~Danilov, Novel photo-detectors and photo-detector systems, Nucl. Instrum.
  Meth. A604~(1) (2009) 183 -- 189.
\newblock \href {http://dx.doi.org/https://doi.org/10.1016/j.nima.2009.01.208}
  {\path{doi:https://doi.org/10.1016/j.nima.2009.01.208}}.

\bibitem{BOHN2009722}
P.~Bohn, A.~Clough, E.~Hazen, A.~Heering, J.~Rohlf, J.~Freeman, S.~Los,
  E.~Cascio, S.~Kuleshov, {\relax Yu}.~Musienko, C.~Piemonte, Radiation damage
  studies of silicon photomultipliers, Nucl. Instrum. Meth. A598~(3) (2009) 722
  -- 736.
\newblock \href {http://dx.doi.org/https://doi.org/10.1016/j.nima.2008.10.027}
  {\path{doi:https://doi.org/10.1016/j.nima.2008.10.027}}.

\bibitem{LI201663}
Z.~Li, Y.~Xu, C.~Liu, Y.~Gu, F.~Xie, Y.~Li, H.~Hu, X.~Zhou, X.~Lu, X.~Li,
  S.~Zhang, Z.~Chang, J.~Zhang, Z.~Xu, Y.~Zhang, J.~Zhao, Characterization of
  radiation damage caused by 23mev protons in multi-pixel photon counter
  (mppc), Nucl. Instrum. Meth. A822 (2016) 63 -- 70.
\newblock \href {http://dx.doi.org/https://doi.org/10.1016/j.nima.2016.03.092}
  {\path{doi:https://doi.org/10.1016/j.nima.2016.03.092}}.

\bibitem{MUSIENKO200987}
{\relax Yu}.~Musienko, D.~Renker, Z.~Charifoulline, K.~Deiters, S.~Reucroft,
  J.~Swain, {Study of radiation damage induced by 82 MeV protons on multi-pixel
  Geiger-mode avalanche photodiodes}, Nucl. Instrum. Meth. A610~(1) (2009) 87
  -- 92.

\bibitem{Chmill:2016msk}
V.~Chmill, E.~Garutti, R.~Klanner, M.~Nitschke, J.~Schwandt, {Study of the
  breakdown voltage of SiPMs}, Nucl. Instrum. Meth. A845 (2017) 56--59.
\newblock \href {http://arxiv.org/abs/1605.01692} {\path{arXiv:1605.01692}},
  \href {http://dx.doi.org/10.1016/j.nima.2016.04.047}
  {\path{doi:10.1016/j.nima.2016.04.047}}.

\bibitem{Klanner:ICASIPM}
S.~Cerioli, E.~Garutti, R.~Klanner, D.~Lomidze, S.~Martens, J.~Schwandt,
  M.~Zvolsky, \href{http://www.icasipm.physics.gatech.edu/}{Characterisation of
  radiation-damaged sipms}, International Conference on the Advancement of
  Silicon Photomultipliers, ICASiPM 2018.
\newline\urlprefix\url{http://www.icasipm.physics.gatech.edu/}

\bibitem{Musienko:2000im}
{\relax Yu}.~Musienko, S.~Reucroft, J.~Swain, {A simple model of EG\&G reverse
  reach-through APDs}, Nucl. Instrum. Meth. A442 (2000) 179--186.
\newblock \href {http://dx.doi.org/10.1016/S0168-9002(99)01218-8}
  {\path{doi:10.1016/S0168-9002(99)01218-8}}.

\bibitem{Angelone:2010mg}
M.~Angelone, M.~Pillon, R.~Faccini, D.~Pinci, W.~Baldini, R.~Calabrese,
  G.~Cibinetto, A.~C. Ramusino, R.~Malaguti, M.~Pozzati, {Silicon
  Photo-Multiplier radiation hardness tests with a beam controlled neutron
  source}, Nucl. Instrum. Meth. A623 (2010) 921--926.
\newblock \href {http://arxiv.org/abs/1002.3480} {\path{arXiv:1002.3480}},
  \href {http://dx.doi.org/10.1016/j.nima.2010.07.057}
  {\path{doi:10.1016/j.nima.2010.07.057}}.

\bibitem{Gotti2018:arxiv}
M.~Calvi, P.~Carniti, C.~Gotti, C.~Matteuzzi, G.~Pessina, {Single photon
  detection with SiPMs irradiated up to $10^{14}$ cm$^{-2}$ 1-MeV-equivalent
  neutron fluence }\href {http://arxiv.org/abs/1805.07154}
  {\path{arXiv:1805.07154}}.

\bibitem{Chatrchyan:2008aa}
S.~Chatrchyan, et~al., {The CMS Experiment at the CERN LHC}, JINST 3 (2008)
  S08004.
\newblock \href {http://dx.doi.org/10.1088/1748-0221/3/08/S08004}
  {\path{doi:10.1088/1748-0221/3/08/S08004}}.

\bibitem{Sze2006}
S.~Sze, K.~N. Kwok, Physics of Semiconductor Devices, 2006.
\newblock \href {http://dx.doi.org/10.1002/0470068329}
  {\path{doi:10.1002/0470068329}}.

\bibitem{ANTUNOVIC2005379}
Z.~Antunovic, I.~Britvitch, K.~Deiters, N.~Godinovic, Q.~Ingram, A.~Kuznetsov,
  {\relax Yu}.~Musienko, I.~Puljak, D.~Renker, S.~Reucroft, R.~Rusack,
  T.~Sakhelashvili, A.~Singovski, I.~Soric, J.~Swain, Radiation hard avalanche
  photodiodes for the cms detector, Nucl. Instrum. Meth. A537~(1) (2005) 379 --
  382.
\newblock \href {http://dx.doi.org/https://doi.org/10.1016/j.nima.2004.08.047}
  {\path{doi:https://doi.org/10.1016/j.nima.2004.08.047}}.

\bibitem{Kirn:1997}
T.~Kirn, et~al., {Wavelength dependence of avalanche photodiode (APD)
  parameters}, Nucl. Instrum. Meth. A387 (1997) 202.

\bibitem{BACCARO1999206}
S.~Baccaro, J.~Bateman, F.~Cavallari, V.~Da~Ponte, K.~Deiters, P.~Denes,
  M.~Diemoz, T.~Kirn, A.~Lintern, E.~Longo, M.~Montecchi, {\relax
  Yu}.~Musienko, J.~Pansart, D.~Renker, S.~Reucroft, G.~Rosi, R.~Rusack,
  D.~Ruuska, R.~Stephenson, M.~Torbet, Radiation damage effect on avalanche
  photodiodes, Nucl. Instrum. Meth. A426~(1) (1999) 206 -- 211.
\newblock \href
  {http://dx.doi.org/https://doi.org/10.1016/S0168-9002(98)01493-4}
  {\path{doi:https://doi.org/10.1016/S0168-9002(98)01493-4}}.

\bibitem{McIntyre:1966}
R.~McIntyre, Multiplication noise in uniform avalanche diodes, IEEE Trans.
  Electron. Dev. 13 (1966) 164--168.

\bibitem{Musienko:2000jj}
{\relax Yu}.~Musienko, S.~Reucroft, D.~Ruuska, J.~Swain, {Studies of neutron
  irradiation of avalanche photodiodes using Cf-252}, Nucl. Instrum. Meth. A447
  (2000) 437--458.
\newblock \href {http://dx.doi.org/10.1016/S0168-9002(99)01019-0}
  {\path{doi:10.1016/S0168-9002(99)01019-0}}.

\bibitem{Musienko:2015fiz}
{\relax Yu}.~Musienko, A.~Karneyeu, {Investigation of avalanche photodiodes
  after irradiation with neutrons up to 5 x 10$^{14}$ n/cm$^2$}, PoS
  PhotoDet2015 (2016) 073.

\bibitem{CMSECAL:1997}
\href{https://cds.cern.ch/record/349375}{{The CMS electromagnetic calorimeter
  project: Technical Design Report}}, Technical Design Report CMS, CERN,
  Geneva, 1997.
\newline\urlprefix\url{https://cds.cern.ch/record/349375}

\bibitem{Vinogradov:2011vr}
S.~Vinogradov, {Analytical models of probability distribution and excess noise
  factor of Solid State Photomultiplier signals with crosstalk}, Nucl. Instrum.
  Meth. A695 (2012) 247--251.
\newblock \href {http://arxiv.org/abs/1109.2014} {\path{arXiv:1109.2014}},
  \href {http://dx.doi.org/10.1016/j.nima.2011.11.086}
  {\path{doi:10.1016/j.nima.2011.11.086}}.

\bibitem{Arosio:2015sra}
V.~Arosio, M.~Caccia, V.~Chmill, A.~Ebolese, M.~Locatelli, A.~Martemiyanov,
  C.~Mattone, R.~Santoro, C.~Tintori, {Development of a Silicon Photomultiplier
  toolkit for science and education}, JINST 10~(07) (2015) C07012.
\newblock \href {http://dx.doi.org/10.1088/1748-0221/10/07/C07012}
  {\path{doi:10.1088/1748-0221/10/07/C07012}}.

\bibitem{Chmill:2016ghf}
V.~Chmill, E.~Garutti, R.~Klanner, M.~Nitschke, J.~Schwandt, {On the
  characterisation of SiPMs from pulse-height spectra}, Nucl. Instrum. Meth.
  A854 (2017) 70--81.
\newblock \href {http://arxiv.org/abs/1609.01181} {\path{arXiv:1609.01181}},
  \href {http://dx.doi.org/10.1016/j.nima.2017.02.049}
  {\path{doi:10.1016/j.nima.2017.02.049}}.

\end{thebibliography}






\end{document}